\documentclass[reprint,
 superscriptaddress,
nofootinbib,
amsmath,amssymb,
aps,
pra,
]{revtex4-2}

\usepackage[utf8]{inputenc} 
\usepackage[T1]{fontenc}    
\usepackage{hyperref}       
\usepackage{url}            
\usepackage{booktabs}       
\usepackage{amsfonts}       
\usepackage{nicefrac}       
\usepackage{microtype}      
\usepackage{graphicx}
\usepackage{xcolor, etoolbox}
\usepackage{amsmath,amsfonts,amsthm} 
\usepackage{mathtools}

\usepackage{tikz}
\usetikzlibrary{positioning,chains}
\usepackage{listofitems} 
\usetikzlibrary{decorations.pathreplacing,calligraphy}
\def\nstyle{int(\lay<\Nnodlen?min(2,\lay):3)} 
\tikzstyle{node}=[very thick,circle,draw=white!60!black,minimum size=22,inner sep=0.5,outer sep=0.6]
\tikzstyle{connect}=[-stealth,thick,blue!40!black,shorten >=1]
\tikzset{ 
  node 1/.style={node,blue!80!black,draw=white!40!black,fill=white!90!black}}
  
\usepackage{ragged2e}
\usepackage{braket}      
\usepackage[english]{babel} 
\usepackage{enumitem}
\usepackage{xspace}
\usepackage{dsfont}
\usepackage{bm}
\usepackage{bbm}
\usepackage{dcolumn}
\usepackage{mathrsfs}

\renewcommand{\vec}[1]{\boldsymbol{#1}}

\hypersetup{
    colorlinks,
    linkcolor=[rgb]{0.206756, 0.371758, 0.553117},
    urlcolor=[rgb]{0.120565, 0.596422, 0.543611},
    citecolor=[rgb]{0.8784313725490196, 0.403921568627451, 0.4549019607843137},
    breaklinks=true 
}

\begin{document}

\title{Learning to detect optical nonclassicality}

\author{Martina Jung}
    \email{martina.jung@uni-jena.de}
    \affiliation{Institute of Condensed Matter Theory and Optics, Friedrich-Schiller-University Jena, Max-Wien-Platz 1, 07743 Jena, Germany}
\author{Suchitra Krishnaswamy}

        \affiliation{Department of Physics, Paderborn University, Warburger Strasse 100, Paderborn 33098, Germany}
    \affiliation{Institute for Photonic Quantum Systems (PhoQS), Paderborn University, Warburger Strasse 100, Paderborn 33098, Germany}
\author{Timon Schapeler}
    \affiliation{Department of Physics, Paderborn University, Warburger Strasse 100, Paderborn 33098, Germany}
    \affiliation{Institute for Photonic Quantum Systems (PhoQS), Paderborn University, Warburger Strasse 100, Paderborn 33098, Germany}
\author{Annabelle Bohrdt}
    \affiliation{Department of Physics and Arnold Sommerfeld Center for Theoretical Physics (ASC),
Ludwig-Maximilians-Universität München, Theresienstraße 37, 80333 München, Germany}
    \affiliation{Munich Center for Quantum Science and Technology (MCQST), Schellingstraße 4, 80799 München, Germany}   
\author{Tim J. Bartley}
    \affiliation{Department of Physics, Paderborn University, Warburger Strasse 100, Paderborn 33098, Germany}
    \affiliation{Institute for Photonic Quantum Systems (PhoQS), Paderborn University, Warburger Strasse 100, Paderborn 33098, Germany}
\author{Jan Sperling}
    \affiliation{Department of Physics, Paderborn University, Warburger Strasse 100, Paderborn 33098, Germany}
    \affiliation{Institute for Photonic Quantum Systems (PhoQS), Paderborn University, Warburger Strasse 100, Paderborn 33098, Germany}
\author{Martin G\"{a}rttner}
    \email{martin.gaerttner@uni-jena.de}
    \affiliation{Institute of Condensed Matter Theory and Optics, Friedrich-Schiller-University Jena, Max-Wien-Platz 1, 07743 Jena, Germany}

\begin{abstract}
Nonclassicality, defined in the quantum optical sense, serves as a resource for photon-based quantum technologies. Therefore, certifying the nonclassicality of a quantum state is crucial for gauging its potential for quantum advantage. However, traditional nonclassicality witnesses that assume perfect knowledge of the witness observables often fail in realistic scenarios with limited statistics and finite-resolution photon detectors. Furthermore, these witnesses do not exploit the fact that certain states are unlikely to be observed in a given experiment.
Here, we train a variational model to distinguish classical from nonclassical states using finitely many measurement samples of multimode quantum states that are probed with different photon-number-resolving detection schemes. The learned decision rule is then an indicator of nonclassicality, tailored to a given set of physically relevant states. Our approach is both data-driven and interpretable in the sense that the learned analytical decision rule can be extracted. Training the model on experimental data measured with (i) a superconducting nanowire single-photon detector and (ii) a time-bin multiplexing detection scheme demonstrates the versatility of the approach, paving the way for efficient nonclassicality detection.
\end{abstract}

\maketitle

\begin{figure*}
    \includegraphics[width=0.8\textwidth]{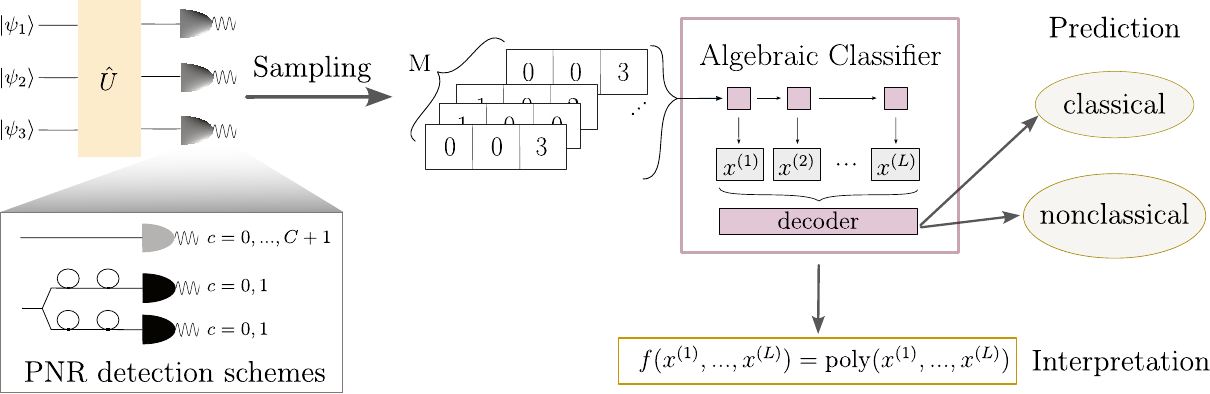}
    \caption{Conceptualization of the classification task. A multimode quantum state is measured with a specific detection scheme (here, the inset depicts a finite PNR detector and a time-bin multiplexing scheme) and stored with the correct binary label: classical or nonclassical. The set of $M$ measurements is input to the algebraic classifier that learns to assign the correct label to the data. First, the encoder (violet boxes at the top) learns to extract relevant moments from the data, then the decoder constructs a polynomial decision rule from these moments. After training, the model is capable of predicting whether a state is nonclassical. Additionally, extracting the learned decision rule allows one to judge and interpret its reliability.}
    \label{fig: visual abstract}
\end{figure*}

\section{Introduction}
While the first universal and fault-tolerant quantum computer is yet to be realized, demonstrating quantum advantage with specialized devices has become its own field of research. In this context, optical nonclassicality is a key resource enabling quantum technologies such as quantum communication \cite{smith_quantum_2011,lercher_standard_2013}, quantum metrology \cite{sahota_quantum_2015,friis_heisenberg_2015, ge_distributed_2018, kwon_nonclassicality_2019} and continuous-variable quantum key distribution \cite{borelli_quantum_2016,anka_introductory_2026}. 
Moreover, it is well known that a nonclassical state produces entanglement at a beam splitter \cite{killoran_converting_2016,vogel_unified_2014,streltsov_colloquium_2017}. Consequently, nonclassicality is also a prerequisite for quantum computing \cite{raussendorf_one-way_2001} and quantum teleportation \cite{bennett_teleporting_1993}. 

Conceptually, nonclassicality is defined on the basis of the Glauber-Sudarshan $P$ representation \cite{sudarshan_equivalence_1963,glauber_coherent_1963}, which allows to expand each state in the coherent state basis $\hat{\rho} = \int P(\vec{\alpha}, \vec{\alpha}^*) |\vec{\alpha}\rangle\langle\vec{\alpha}|\mathrm{d}^2\vec{\alpha}$. Due to the overlapping nature of the coherent states, the $P$ representation of a state is not unique \cite{drummond_generalised_1980}. If $\hat{\rho}$ can be represented with a nonnegative function $P$, the state is called classical. Conversely, a state is nonclassical if its $P$ representation cannot be interpreted as a probability distribution.

Since the identification of nonclassicality as a resource, a range of methods for experimentally verifying the nonclassicality of a quantum state has been developed. Early works are based on the photon-number distribution and moments thereof \cite{mandel_sub-poissonian_1979,agarwal_nonclassical_1992, klyshko_observable_1996, innocenti_nonclassicality_2022}.
However, state-of-the-art photon-number-resolving (PNR) detection schemes resort to multiplexing with click detectors \cite{paul_photon_1996, achilles_fiber-assisted_2003, fitch_photon-number_2003, bartley_superconducting_2023}, as well as the finite PNR capacity of superconducting nanowire single-photon detectors (SNSPDs) \cite{cahall_multi-photon_2017, sauer_resolving_2023, schapeler_electrical_2024} or transition edge sensors \cite{lucia_transition_2024}. These techniques have motivated the development of tailored nonclassicality criteria for multiplexing schemes \cite{sperling_true_2012, sperling_sub-binomial_2012, sperling_correlation_2013, filip_hierarchy_2013, luis_nonclassicality_2015, sperling_detector-independent_2017, bohmann_detection-device-independent_2019}.
Later works established nonclassicality witnesses based on the distinguishability from a state with a positive Wigner function \cite{mari_directly_2011}, phase-space distributions \cite{bohmann_probing_2020, biagi_experimental_2021}, the characteristic function \cite{richter_nonclassicality_2002,ryl_unified_2015}, and semidefinite programming \cite{korbicz_hilberts_2005,ohst_revealing_2025}. The major drawback of these witnesses lies in the necessity to have full knowledge about the witness observables. In realistic experimental settings, estimates of observables suffer from statistical noise and detector inefficiencies, rendering the evaluation of witness observables costly or even infeasible.
Additionally, these traditional witnesses do not take into account that specific states are more likely to be observed in a particular experiment than others.

Here we argue that in realistic application scenarios, the most useful tool is not a (sufficient) witness that assumes perfect knowledge about some observables but a sample-efficient indicator that allows one to judge whether a state is potentially nonclassical. Consider a situation in which we have access to labeled data, originating from either (i) a test run of a trusted experimental device or (ii) a simulation that is performed with a noise-free unitary. This data then serves as a baseline of what one agrees on to acknowledge as classical and nonclassical in the context of the experiment. 
Learning a decision rule that identifies the presence or absence of nonclassical features enables real-time monitoring of the experimental performance in preparing nonclassical resource states, which may deteriorate due to parameter drifts or increased thermal noise.
The sample efficiency of the nonclassicality indicator is essential for ensuring the applicability to situations where data acquisition is limited and costly.

Employing machine learning methods for efficiently certifying the nonclassicality of a state, while bypassing quantum state tomography, has been previously suggested \cite{ahmed_classification_2021,gebhart_neural-network_2020,gebhart_identifying_2021, machulka_revealing_2024}. Gebhart et al.\ designed a neural network that learns to identify whether a state is nonclassical based on photon-count probabilities estimated with finite statistics \cite{gebhart_neural-network_2020}. Later, the architecture was also applied to experimental homodyne detection data \cite{gebhart_identifying_2021}. Comparing the model's prediction with known nonclassicality signatures of the states allowed for a post-hoc interpretation of the learned features. However, the neural network itself remained a black box. Again, it must be noted that the prediction of a model, trained on informationally-incomplete data, can only ever be interpreted as an indicator for nonclassicality. Therefore, we will use the term criterion for a non-sufficient nonclassicality indicator that was learned by a variational model.

In this contribution, we present the algebraic classifier (AlCla), a variational model, which is trained via supervised learning to distinguish classical from nonclassical states. By leveraging prior knowledge about the families of states that are realistically observed, the model learns a decision rule that is optimized for a given experiment and detection scheme. The objective is not to learn a nonclassicality criterion that generalizes to nonclassical states outside of the training data distribution, but a criterion that interpolates to states within the same family but with different parameters compared to those seen during training.
Figure \ref{fig: visual abstract} schematically depicts the data acquisition and the training process. The algebraic classifier is interpretable in the sense that the learned decision rule can be extracted once the model has been trained. It is important to stress that the transparency of the model does not equal instant interpretability; that is, an exact formula is rigorous but might lack intuition of the underlying physics \cite{wetzel_interpretable_2025}.

The paper is organized as follows: In Sec.~\ref{sec:preliminaries}, we give an overview of moment-based and probability-based nonclassicality witnesses. We start by discussing witnesses based on measurements in the photon-number basis and then turn to the less studied case of click statistics, see also \cite{krishnaswamy2026observable}.  These witnesses will serve as a reference to which we compare our approach.
In Sec.~\ref{sec:architecture}, we introduce the AlCla and discuss how higher-order moments are hierarchically constructed in the encoder. In Sec.~\ref{sec:results}, we present our results showing that the AlCla is capable of both rediscovering and outperforming traditional nonclassicality witnesses. We train the model on three single-mode datasets simulated with (i) a perfect PNR detector, (ii) an SNSPD with finite photon-number resolution \cite{schapeler_practical_2025}, and (iii) a time-bin multiplexing scheme including two click SNSPDs with four time-bins each \cite{schapeler_information_2022}. Two of the datasets were simulated using experimentally reconstructed positive operator-valued measures (POVMs) that fully characterize the detectors \cite{luis_complete_1999, fiurasek_maximum-likelihood_2001,lundeen_tomography_2009, feito_measuring_2009}. Furthermore, we demonstrate the model's applicability to multimode systems by training it on a 6-mode dataset, simulated with a fixed random unitary and measured with a PNR SNSPD.
Finally, in Sec.~\ref{sec:conclusions}, we give an outlook on possible future research directions.

\section{Preliminaries}
\label{sec:preliminaries}
A necessary and sufficient condition for a state $\hat{\rho}$ to be nonclassical is that there exists an operator $\hat{f} = \hat{f}(\hat{a}^\dagger, \hat{a})$ such that the expectation value $\langle: \hat{f}^\dagger\hat{f}:\rangle = \int |f(\vec{\alpha},\vec{\alpha}^*)|^2 P(\vec{\alpha},\vec{\alpha}^*)\textnormal{d}^2\vec{\alpha} $ is negative \cite{shchukin_nonclassicality_2005}. Here, the colons $:\cdot:$ indicate normal ordering.
Approaches based on semidefinite programming (SDP) \cite{korbicz_hilberts_2005, ohst_revealing_2025} recast the question of whether a state $\hat{\rho}$ is nonclassical into a minimization problem over the set of normal ordered observables $\hat{W}$: A state $\hat{\rho}$ is nonclassical if and only if
\begin{equation}
    \min_{\hat{W}} \left\lbrace \langle \hat{W}\rangle_{\hat{\rho}}: \langle \alpha| \hat{W} |\alpha\rangle \geq 0\right\rbrace <0.
\end{equation}
Here, $\lbrace|\alpha\rangle\rbrace$ represents the set of coherent states. 

In the context of an experimental setup, it is important to note that not all possible operators $\hat{W}$ will be accessible. Consequently, the search space is restricted to the space of available operators, denoted by $\mathcal{F}$. The criterion $\exists f\in \mathcal{F}:\langle :\hat{f}^\dagger\hat{f}:\rangle<0$ then becomes a sufficient witness for nonclassicality. Consequently, there exist nonclassical states which cannot be witnessed by any $\hat{f}\in \mathcal{F}$. In this work, we will focus on operators which are functions of the photon-number operator.
First, we give an overview of sufficient nonclassicality criteria based on photon-number measurements. Then, we discuss the less investigated but more realistic case of multiplexing detection schemes. Although the next two subsections discuss single-mode nonclassicality criteria, all --- except the Klyshko criterion \cite{klyshko_observable_1996} --- can be generalized to multiple modes, as outlined in Appendix~\ref{sec: derivation of 3rd order mom} and~\ref{sec: derivation of gen. Klyshko}.

\subsection{Photon-number detection}
In this section, we review nonclassicality witnesses accessible in experiments with PNR detectors. For finite detection efficiency $\eta<1$ and dark counts $\nu>0$, the photon-counting statistics is given by \cite{mandel_sub-poissonian_1979}
\begin{equation}
    p_n = \left\langle : \frac{(\eta\hat{n}+\nu)^n}{n!} e^{-(\eta\hat{n}+\nu)} :\right\rangle.
\end{equation}
where $\hat{n} = \hat{a}^\dagger\hat{a}$ is the photon-number operator.

\subsubsection{Moment-based witnesses}
The first experimentally accessible nonclassicality witness was proposed by Mandel \cite{mandel_sub-poissonian_1979} who found that the parameter $Q = \frac{\langle \hat{n}^2\rangle - \langle \hat{n}\rangle^2}{\langle \hat{n}\rangle}-1 \overset{\mathrm{cl.}}{\geq} 0$ is nonnegative for all classical states. This Mandel $Q$ parameter has a clear physical interpretation as it quantifies the deviation of a state's photon-number distribution from Poissonian statistics. Coherent states result in Poissonian statistics with $Q=0$, while number states realize the minimum possible value of $Q=-1$ due to their singular photon-number distribution.

Agarwal et al.\ \cite{agarwal_nonclassical_1992} found that the Mandel $Q$ parameter is a special instance of a more general hierarchy of sufficient nonclassicality criteria that are based on a matrix of normal-ordered moments
\begin{equation}
    \begin{pmatrix}
    1 &  \langle \hat{n}\rangle &  \langle :\hat{n}^2:\rangle & ...\\
     \langle \hat{n}\rangle &  \langle :\hat{n}^2:\rangle &  \\
      \langle :\hat{n}^2:\rangle & & \ddots \\
      \vdots
\end{pmatrix}
\end{equation}
which is positive semi-definite for classical states. Applying Sylvester's criterion \cite{horn_matrix_1985}, which states that a matrix is positive semi-definite iff all of its principal minors are positive semi-definite, yields a hierarchy of moment-based witnesses. While the first leading principal minor gives a second-order witness equivalent to the Mandel $Q$ parameter, the next leading principal minor already gives a fourth-order witness. In order to derive a third-order witness, Krishnaswamy et al.\ \cite{krishnaswamy2026observable} developed an alternative approach to construct the matrix of moments. To ensure that this work is self-contained, the derivation of the witness and its generalization to multiple modes are presented in Appendix~\ref{sec: derivation of 3rd order mom}. The first leading principal minor then gives the following third-order witness:
\begin{equation}
     Q_3 = \langle \hat{n}\rangle\langle\hat{n}^3\rangle -\langle\hat{n}^2 \rangle^2 +  \langle \hat{n}^2\rangle\langle\hat{n}\rangle + \langle\hat{n}\rangle^2 \overset{\mathrm{cl.}}{\geq}0.
\end{equation}

\subsubsection{Probability-based witnesses}
Complementary to the moment-based witnesses described above, probability-based approaches consider the probability of detecting a certain number of photons.
The most prominent witness goes back to Klyshko who found that the violation of
\begin{equation}
    \frac{(k+1)p_{k-1}p_{k+1}}{kp_k^2}\overset{\mathrm{cl.}}{\geq}1
\end{equation}
is a sufficient criterion for nonclassicality \cite{klyshko_observable_1996}. One advantage of Klyshko's witness over moment-based witnesses is that the latter give incorrect results for truncated photon-number statistics which appear in the context of a detector with finite photon-number resolution.
By applying standard techniques laid out in \cite{innocenti_nonclassicality_2022}, Krishnaswamy et al.\ generalized Klyshko's criterion and derived the following probability-based witness \cite{krishnaswamy2026observable} (for a summary of the derivation, see Appendix~\ref{sec: derivation of gen. Klyshko}):
\begin{align}
    \begin{split}
    0 \overset{\mathrm{cl.}}{\leq} & (M_{jk})_{j,k} = \left(\binom{j+k}{j} p_{j+k} \right)_{j,k}, \\
    & j,k \in \mathbb{N} \textnormal{ or } j,k\in\frac{1}{2}\mathbb{N}\setminus \mathbb{N}.
    \end{split}
\end{align}
Depending on whether $j,k$ are chosen to be integers or half-integers, the witness is sensitive to even or odd photon-number parity. Note that the binomial coefficient can be applied to real numbers using the gamma function $\binom{k}{y} = \frac{k!}{\Gamma(y+1)\Gamma(k-y+1)}$. We will refer to this witness as the generalized Klyshko witness. 
In contrast to the original Klyshko criterion, it is possible to generalize the witness to multiple modes. However, for limited statistics, the statistical errors become exceedingly large due to the exponential growth of the multimode Hilbert space.

\subsection{Multiplexing detection schemes}
Whilst the realization of PNR detectors represents a highly active area of research, the focus has partly shifted to multiplexing detection schemes employing available threshold detectors. This scheme can be realized with spatial \cite{paul_photon_1996} or temporal \cite{achilles_fiber-assisted_2003} multiplexing, or a combination of both \cite{harder_single-mode_2016,eaton_resolution_2023}. The underlying idea is to split an incident beam of photons into multiple modes such that the chance of multiple photons populating the same mode is reduced. Compared to spatial multiplexing schemes, time-binning offers the advantage of being resource-efficient because the number of detectors required is independent of the number of temporal modes. It is important to note that the click statistics is fundamentally different from the underlying true photon-number distribution. Hence, traditional nonclassicality witnesses have to be adapted.

Following reference \cite{sperling_true_2012}, the POVM of the no-click event in a multiplexing scheme using threshold detectors and $N$ (temporal or spatial) modes is given by $\hat{\pi}_0 = e^{-\eta\frac{\hat{n}}{N}+\nu}$. The POVM of $k$ clicks is then given by $\hat{\Pi}_k = \binom{N}{k} \left(1-\hat{\pi}_0\right)^k \hat{\pi}_0^{N-k}$. Taking the normal-ordered expectation value of this expression yields the probability for measuring $k$ clicks
\begin{equation}
    c_k = \left\langle: \binom{N}{k} \left(1-e^{-(\eta\frac{\hat{n}}{N}+\nu)}\right)^{k}\left(e^{-(\eta\frac{\hat{n}}{N}+\nu)}\right)^{N-k} :\right\rangle.
    \label{eq: click distribution probs}
\end{equation}

\subsubsection{Moment-based witnesses}
In a multiplexing scheme with threshold detectors, coherent states do not produce a Poissonian, but a binomial distribution. Thus, nonclassicality can be witnessed by the negativity of the binomial parameter \cite{sperling_sub-binomial_2012, lee_sub-poisson-binomial_2016,bartley_directly_2013}
\begin{equation}
    Q_B = \langle \hat{c}^2\rangle - \frac{N-1}{N}\langle \hat{c}\rangle^2-\langle \hat{c}\rangle \overset{\mathrm{cl.}}{\geq}0.
    \label{eq: def of QB}
\end{equation}
Here, $N$ is the total number of bins in the detection scheme and $\langle \hat{c}^{m}\rangle = \sum_{k=0}^N k^m c_k$ are the moments of the click distribution. In Sec.~\ref{time-bin multiplexing scheme} we will consider a detection scheme with two threshold detectors and four time-bins per detector, resulting in $N=8$ bins in total.

Analogous to the third-order moment-based witness for PNR detectors, one can formulate the following third-order witness (derivation in Appendix~\ref{sec: derivation of 3rd order mom}):
\begin{equation}
\begin{split}
    0\overset{\mathrm{cl.}}{\leq} Q_{B,3} &= \langle \hat{c}^3\rangle\langle \hat{c}\rangle - \frac{N-2}{N-1}\langle\hat{c}^2\rangle^2 - \frac{N+1}{N-1}\langle \hat{c}^2\rangle\langle\hat{c}\rangle \\
    + &\frac{N}{N-1} \langle \hat{c}\rangle^2.
\end{split}
\end{equation}

\subsubsection{Probability-based witnesses}
Alternatively, the click distribution in Eq.~\eqref{eq: click distribution probs} itself gives rise to the following probability-based criterion
\begin{equation}
    \begin{split}
    0 \overset{\mathrm{cl.}}{\leq} M_{j,k} = \left( \frac{c_{k+j}}{\binom{N}{j+k}} \right)_{j,k}, j,k \in\mathbb{N} \textnormal{ or } j,k\in\frac{1}{2}\mathbb{N}\setminus \mathbb{N}
    \end{split}
\end{equation}
that will be called generalized Klyshko witness as in the case of a PNR detector.

\section{The algebraic classifier}
\label{sec:architecture}
In this section, we present the algebraic classifier (\mbox{AlCla}), a variational model that is trained via supervised learning to distinguish classical from nonclassical states. Our model is designed to be sample-efficient while learning a decision rule whose structure is similar to the criteria derived from the matrix of moments. Hence, the key point distinguishing our model from traditional moment-based witnesses is the lack of sufficiency. Instead of a witness, the model learns an indicator of nonclassicality. This criterion can be interpreted as a decision boundary that separates classical from nonclassical states represented by points in a high-dimensional space. Specifically, we envision different families of states to form point clouds in the space spanned by the statistical moments $\langle\hat{n}_1\rangle, \langle\hat{n}_1\hat{n}_2\rangle, \langle\hat{n}_1^2\rangle$, etc. 

\subsection{Architecture}
\begin{figure}
    \includegraphics[width=0.9\linewidth]{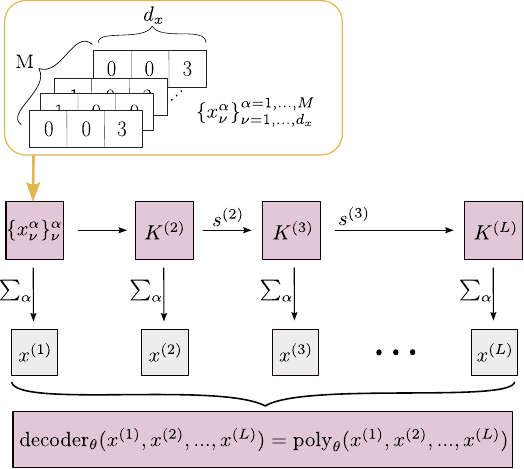}
    \caption{Structure of the algebraic classifier. The set of sampled photon numbers/clicks $\left \lbrace x_{\nu}^{\alpha}\right\rbrace$ is input to the encoder that learns to extract relevant moments up to order $L$. These weighted sums of moments are stored in $\lbrace \vec{x}^{(i)}\rbrace_{i=1,...,L}$ and input to the decoder. The decoder constructs a polynomial of order $L$ and optimizes its coefficients.}
    \label{tikzfig:interpretable decoder schematic}
\end{figure}

The AlCla has an encoder-decoder structure which is schematically presented in Fig.~\ref{tikzfig:interpretable decoder schematic}. While the encoder extracts relevant moments from the photon/click measurements, the decoder learns to combine this information in order to make a prediction about the nonclassicality of the state.
The input of the model is the set of measurements $\lbrace x_\nu^\alpha\rbrace_{\nu=1,...,d_x}^{\alpha = 1,...,M}$ with $x_\nu^\alpha$ describing the number of photons/clicks detected in mode $\nu$ in the measurement run $\alpha$. Here, $d_x$ corresponds to the number of modes and $M$ to the total number of measurement samples. The weights $K^{(i)}$ in the encoder determine which inter-mode correlations contribute to the intermediate output $\vec{x}^{(i)}$ of order $i$. The output of the first layer is computed by summing over the physical index $\nu$ and the sample index $\alpha$, thus acknowledging the permutation invariance of the measurement samples
\begin{equation}
    x_\nu^{(2)} = \sum_{\alpha=1}^{M}\sum_{\eta=1}^{d_x} K^{(2)}_{\nu\eta}x_{\nu}^{\alpha}x_{\eta}^\alpha.
    \label{eq: output first layer}
\end{equation}
Here, $K^{(2)}\in \mathbb{R}^{(d_x, d_x)}$ is a learnable matrix. For example, if $K^{(2)}_{\nu\eta} = \delta_{\nu,\eta}$, the $\nu^{th}$ entry of $\vec{x}^{(2)}$ is proportional to $\langle x_\nu^2\rangle$.

In order to hierarchically construct higher-order moments, the input to the next layer of the encoder is
\begin{equation}
    \left(s^{(2)}\right)_{\nu}^{\alpha} := \sum_{\eta=1}^{d_x} K^{(2)}_{\nu\eta}x_{\nu}^\alpha x_{\eta}^\alpha,
\end{equation}
the summand in Eq.~\eqref{eq: output first layer}. The output of the second layer is then given by
\begin{align}
    x_\nu^{(3)} &= \sum_{\alpha=1}^{M}\sum_{\eta=1}^{d_x}K^{(3)}_{\nu\eta}x_{\nu}^{\alpha}\left(s^{(2)}\right)_{\eta}^\alpha\\
    &=\sum_{\alpha=1}^{M}\sum_{\eta=1}^{d_x} K^{(3)}_{\nu\eta} x_{\nu}^{\alpha}\sum_{\mu=1}^{d_x} K^{(2)}_{\eta\mu}x_{\eta}^\alpha x_{\mu}^\alpha \\
    &=\sum_{\alpha=1}^M \underbrace{\sum_{\eta,\mu=1}^{d_x} K^{(3)}_{\nu\eta} K^{(2)}_{\eta\mu}x_{\nu}^{\alpha}x_{\eta}^\alpha x_{\mu}^\alpha}_{=\left(s^{(3)}\right)_{\nu}^\alpha}.
\end{align}
Each entry $x_\nu^{(i)}$ represents a weighted sum of moments of order $i$, associated with mode $\nu$. From this construction, it is evident that higher-order moments strongly depend on which lower-order moments are selected in previous layers. It is this interdependence of the extracted moments that impedes the model from calculating all possible moments, thereby ensuring its capability to generalize to multimode systems. The number of parameters in the encoder scales quadratically with the number of modes. In total, there are $d_x(d_x+1)(L-1)/2$ parameters in the encoder for $d_x$ modes and $L-1$ encoding layers.

The set $\left\lbrace x_\nu^{(i)}\right\rbrace_{\nu=1,...,d_x}^{i=1,...,L}$ is input to the algebraic decoder which constructs a polynomial $f$ of total order $L$, that is, a polynomial which only includes terms of the form $\left(x_{i1}^{(m_1)}\right)^{j_1}\left(x_{i2}^{(m_2)}\right)^{j_2}...$ with $\sum_{l}j_lm_l\leq L$.
The coefficients $\vec{\theta}$ of the polynomial are learnable weights, making the decoder a specialized form of nonlinear regression model. The underlying intuition is that the polynomial resembles a nonclassicality criterion obtained from the matrix of moments: For a given input state, the sign of the polynomial's value determines whether a state is classical or nonclassical. To enhance the separation of the two classes, the output of the polynomial is amplified with a learnable factor $\theta_\mathrm{amplify}$. 
Finally, the function $1-\mathrm{sigmoid}(\cdot)$ converts the sign of the polynomial to a value $y\in[0,1]$. 
With this construction, we opt against a symbolic regression-based approach, as implemented in \cite{kharkov_discovering_2021}, which is more appropriate for scenarios where the analytical formula is expected to be sparse.

To illustrate the construction in the decoder, we consider a model with two encoding layers that extracts up to third-order moments. For a single mode, the constructed polynomial will take the following form:
\begin{align}
    \begin{split}
        f(x^{(1)},x^{(2)},x^{(3)}) &= \theta_1 x^{(1)}+\theta_2(x^{(1)})^2 + \theta_3 (x^{(1)})^3\\
        & + \theta_4 x^{(2)} + \theta_5 x^{(3)} + \theta_6 x^{(1)}x^{(2)} \\
        &+ \theta_7.
    \end{split}
\end{align}
Hence, for $L-1$ encoding layers, the model learns a nonclassicality criterion based on photon-/click-number moments up to order $L$.
For the scenarios considered in this work, we find that going beyond $L=3$ does not lead to a significant improvement of the model's performance. For $L\leq3$, the polynomial consists solely of single, double, and triple terms. Taking into account only up to triple terms for general $L$, we find that the number of parameters in the decoder, denoted by $N_P$, is bounded from above by $1+\frac{d_x\tilde{L}}{36}\left(35+d_x^2\tilde{L}^2\right)$. Here, $\tilde{L}=L\ln(L)+L\gamma$ and $\gamma\approx0.5772$ is the Euler-Mascheroni constant. It is important to note that, although this bound correctly predicts the scaling with $L$ and $d_x$, it is generally not tight; details can be found in Appendix~\ref{sec: details about AlCla}.
Regarding scalability in terms of experimental system size, the relevant question is how $N_P$ scales with the mode number $d_x$ for a fixed number of $L-1$ encoding layers: For $d_x\gg L$, $N_P\sim \mathcal{O}\left(d_x^L\right)$.

As an alternative model, we consider support vector machines (SVM) that represent a geometrically interpretable approach for the classification of multi-class data \cite{burges_tutorial_1998}. When discussing the results for the 6-mode dataset in Sec.~\ref{sec: results 6mode ds}, we will compare the model's findings against a linear SVM. 
Although the kernel trick enables a nonlinear classification boundary, choosing the best kernel for a given problem is still an open question. The key difference between our model and an SVM (with a linear kernel) is that the AlCla allows for nonlinear terms up to a certain order while the SVM's decision rule is linear in its inputs (the moments of the photon-number distribution). Additionally, the AlCla enables the steering of the model's bias towards one of the classes, as discussed in the next subsection. An SVM also allows for weighting elements of different classes differently. However, this requires manual weighting of the individual samples.

\subsection{Training}
The model is trained to minimize the binary cross entropy. In order to control the rate of false positives, i.e.\ the number of classical states incorrectly classified as nonclassical, we add a regularization to the loss function:
\begin{equation}
    \mathcal{L} = -\tilde{y} \log(y) - (1-\tilde{y}) \log(1-y) + \lambda(1-\tilde{y}) |\tilde{y}-y|.
    \label{def: loss function}
\end{equation}
Here, $\tilde{y}$ is the true label (0 for classical, 1 for nonclassical states) and $y$ corresponds to the prediction of the model. Increasing the regularization strength $\lambda$ punishes the model for misclassifying classical states. Consequently, the model's prediction that a state is nonclassical becomes a more reliable indicator of nonclassicality.

To minimize the cost function, we employed the Adam optimizer in conjunction with batch gradient descent and weight clipping for the learnable parameters: $$K^{(i)}\in[-10,10],\; \theta_\mathrm{amplify}\in[1,50], \;\theta_i \in[-10,10].$$

\subsection{Training dataset}
We train the model on datasets measured with different detection schemes, incorporating experimentally retrieved detector characteristics. The composition of each dataset is described in detail in the corresponding subsection in Sec.~\ref{sec:results}. The simulated data, including the amplitudes of the states, are available publicly in the Zenodo repository~\cite{jung2026learning}.
The focus of this work lies on the following Gaussian and non-Gaussian states:
\begin{itemize}
    \item coherent states: $$|\alpha\rangle = e^{-\frac{|\alpha|^2}{2}}\sum_{n=0}^\infty \frac{\alpha^n}{\sqrt{n!}}|n\rangle$$ with $\alpha$ being the amplitude of the coherent state.
    \item mixed coherent states: $$\hat{\rho}_\mathrm{mix} = \frac{|\alpha_1\rangle\langle \alpha_1| +|\alpha_2\rangle\langle \alpha_2|}{2}$$ with $\alpha_1, \alpha_2$ being the amplitudes of the two coherent states.
    \item thermal states: $$\hat{\rho}_\mathrm{th} = \sum_{m=0}^\infty \frac{\bar{n}^m}{(1+\bar{n})^{m+1}}|m\rangle\langle m|$$ with $\bar{n}$ being the mean number of thermal photons.
    \item squeezed vacuum: $$|r\rangle = \frac{1}{\sqrt{\cosh(r)}}\sum_{n=0}^\infty (-\tanh(r))^n \frac{\sqrt{(2n)!}}{2^nn!}|2n\rangle$$ with $r\in\mathbb{R}$ being the squeezing parameter.
    \item single-photon-added thermal states (SPATS): $$\hat{\rho}_\mathrm{SPATS}(\bar{n})=\frac{\hat{a}^\dagger \hat{\rho}_\mathrm{th}(\bar{n})\hat{a}}{\mathrm{tr}(\hat{\rho}_\mathrm{th}(\bar{n})\hat{a}\hat{a}^\dagger )}.$$ The probability of finding $m$ photons is $p_m = m\frac{\bar{n}^{m-1}}{(1+\bar{n})^{m+1}}$, with $\bar{n}$ being the mean number of thermal photons \cite{barnett_statistics_2018}.
    \item single-photon-loss photon-number states (PNS): $$\hat{\rho}_{n} = (1-p_\mathrm{loss})|n\rangle\langle n| + p_\mathrm{loss} |n-1\rangle\langle n-1|.$$
\end{itemize}

\section{Results}
\label{sec:results}
In this section, we discuss the results of our model in learning nonclassicality indicators that are optimized for a given set of states.
We start with the single-mode case and the rediscovery of a traditional witness. Then, we continue with different detection schemes and experimentally reconstructed detection POVMs showing that the model outperforms traditional witnesses. Finally, we test the model's capability to generalize to multimode setups by training on a 6-mode dataset.

Each analysis comprises a brief summary about the dataset used for training. In all cases, $80\%$ of the states were used for training while the remaining states were kept for testing. The model showed equivalent performance on the training and test dataset, demonstrating its interpolating capabilities (for a typical loss curve during training, see Fig.~\ref{fig: typical loss curve} in the Appendix). Consequently, this section only discusses the performance on the training dataset. It is important to note that the objective of the model is not to generalize to unseen states that exhibit different signatures of nonclassicality. Instead, we aim for a model that can identify nonclassical signatures that are expected to appear in a specific experiment when working with particular families of input states.

In all cases --- except for the case $d_x=6, L=3$ ---, the model is under-parameterizing the data. In the case of an over-parameterizing model, we will see that the same performance can be reached with a smaller model ($L=2$) and a higher regularization strength $\lambda$. Hence, our model is not overfitting the data, but learning physical features.

The analysis focuses on the algebraic classifier (AlCla) being trained with one and two encoding layers, corresponding to second- and third-order criteria. An encoder with more than three encoding layers has proven to not increase the total accuracy, i.e.\ the ratio of correctly predicted states, further.
For the first dataset, we use a constant learning rate and dynamically update the best epoch for the evaluation of the model's performance. Otherwise, the model is trained for $900$ epochs with an initial learning rate of $10^{-2}$ and a scheduler that halves the learning rate once the train accuracy reaches a plateau. In this case, the performance of the model is extracted after the last training epoch.

\subsection{Rediscovering and outperforming known nonclassicality witnesses}
In order to rediscover the Mandel $Q$ parameter as the optimal nonclassicality criterion, we train the model on a dataset comprising coherent, mixed coherent and squeezed states, and SPATS. The single-mode states are simulated assuming a noise-free photon-number detector with a resolution of up to 29 photons. The amplitudes of the states were chosen in a lower range so that the truncated tail of the photon-number distribution becomes negligible. Table~\ref{tab: ds 12 train} details the composition of the dataset.
\begin{table}[b]
\caption{\textbf{Composition of the perfect PNR detector dataset.} The amplitudes $r,\bar{n},\alpha$ of the states lie equidistantly in the amplitude range. In total, there are $32$ nonclassical states and $54$ classical states. The cutoff of the detector was set to $29$.}
\label{tab: ds 12 train}
\centering
\begin{tabular}{ccc}
    \toprule\toprule
    species & amplitude range & number of states \\\midrule
    squeezed & $r \in [0.1, 1.2]$ & $12$\\
    SPATS & $\bar{n} \in [0.25,1.2]$ & $20$\\
    coherent & $\alpha \in [0.0, 3.5]$& $36$\\
    mixed coherent & $\alpha\in[0.0,3.5]$ & $18$\\
    \toprule\toprule
\end{tabular}
\end{table}
The model was trained with a constant learning rate of $10^{-2}$ and the epoch used to evaluate the model's performance was selected to be the last epoch that met the following conditions: (i) the accuracy on either the training or the test dataset increased and (ii) the performance of the other dataset remained at least unchanged.
\begin{figure}
    \centering
    \includegraphics[width=\linewidth]{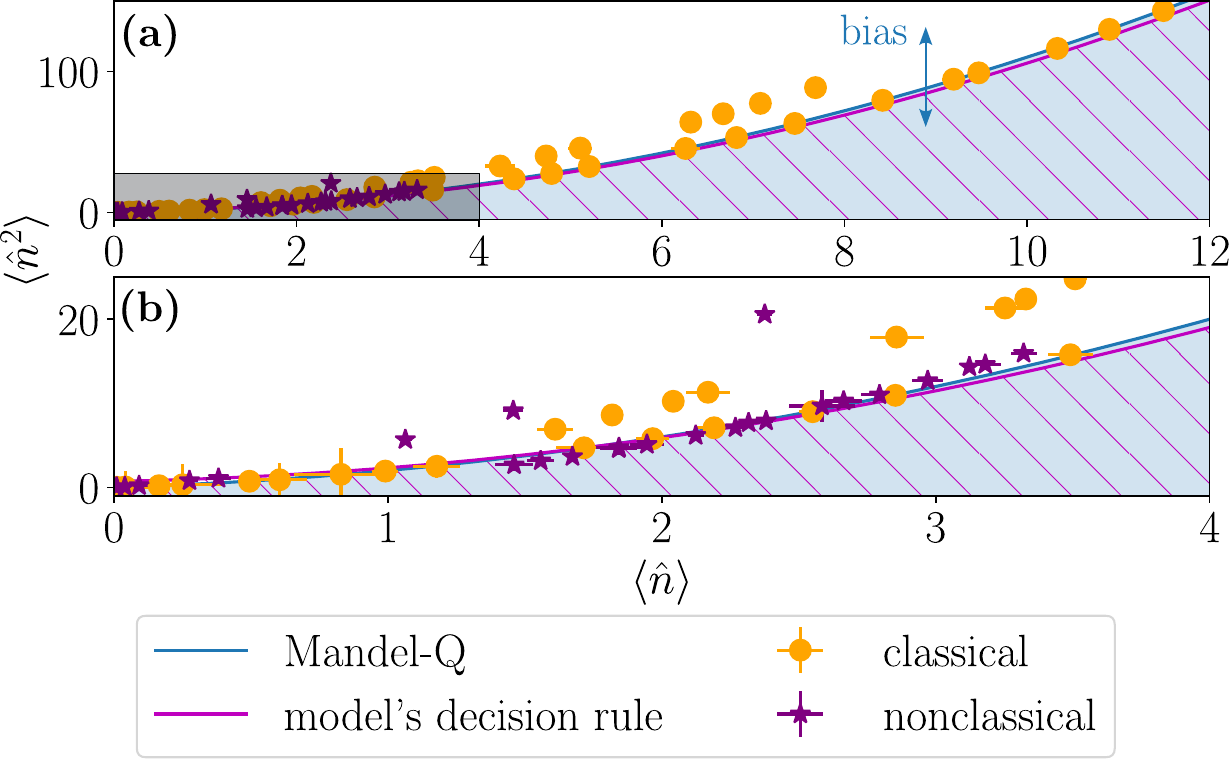}
    \caption{Visualization of the dataset simulated with a perfect PNR detector and $10^3$ samples per state: Orange circles (purple stars) represent classical (nonclassical) states. The shaded blue area visualizes the set of states that exhibit sub-Poissonian statistics. (a) To account for statistical errors of the moments, a bias can be added to the witness. Geometrically, this corresponds to a vertical shift of the decision boundary (blue arrow). 
    The magenta line represents the learned decision rule of the model that is trained with a single encoding layer, a constant learning rate and no regularization $\lambda=0$. The hatched area visualizes the set of states that is identified as nonclassical by the model.
    (b) Zoom of the regime with small $\langle \hat{n}\rangle$ and $\langle\hat{n}^2\rangle$ marked as gray box in (a). Errorbars represent the standard error.}
    \label{fig: Visualization datapoints ds12}
\end{figure}
Figure \ref{fig: Visualization datapoints ds12} visualizes the data points in the space spanned by the moments $\langle\hat{n}\rangle$ and $\langle\hat{n}^2\rangle$ --- which are available to a model with one encoding layer --- for $M=10^3$ measurement samples per state. Panel (b) shows the regime of small $\langle \hat{n}\rangle$ and $\langle\hat{n}^2\rangle$ marked as gray box in (a). Error bars represent the standard error $\sqrt{\sigma/M}$ with $\sigma$ being the variance of the respective moments.
The coherent states are located approximately on the parabola defined by $Q=0$ (blue line). However, due to the limited statistics, some coherent states fall into the blue shaded area, i.e.\ exhibit slightly sub-Poissonian statistics. To account for this statistical error, it is necessary to add a bias to the witness $Q \rightarrow Q+\mathrm{bias}$, which corresponds to a vertical shift of the parabola in the picture. Adding a positive bias will increase the number of correctly identified classical states at the expense of potentially misclassifying some nonclassical states. Thus, adding a bias effectively means classifying states as nonclassical only if they violate the witness by a certain statistical error margin.

When trained with a single encoding layer, a constant learning rate and no regularization $\lambda=0$, the model learns a slight modification of the Mandel $Q$ parameter, represented by the magenta curve. Due to the tilt in the parabola, the model is able to correctly identify larger-amplitude coherent states as classical even though they exhibit a slightly sub-Poissonian statistics (potentially due to finite truncation at $n=29$ photons). Moreover, the modification takes into account that the nonclassical states are concentrated at lower moments $\langle \hat{n}\rangle$. This highlights that the learned decision rule strongly depends on the dataset and the distribution of points in the plane spanned by $\langle\hat{n}\rangle$ and $\langle\hat{n}^2\rangle$.

An ideal classifier would achieve $100\%$ in identifying both classes of states correctly. However, Fig.~\ref{fig: Visualization datapoints ds12} demonstrates that the two classes cannot be separated by a simple parabolic decision boundary when only up to second-order moments are available. In reality, any classifier must find a compromise between achieving a high accuracy on classical or nonclassical states. The AlCla's compromise can be implicitly tuned by the regularization $\lambda$, defined in Eq.~\eqref{def: loss function}, whereas traditional witnesses can be directly modified with an additional bias term. Increasing these two variables parametrizes the AlCla's and the traditional witnesses' trade-off. 
\begin{figure}
    \includegraphics[width=\linewidth]{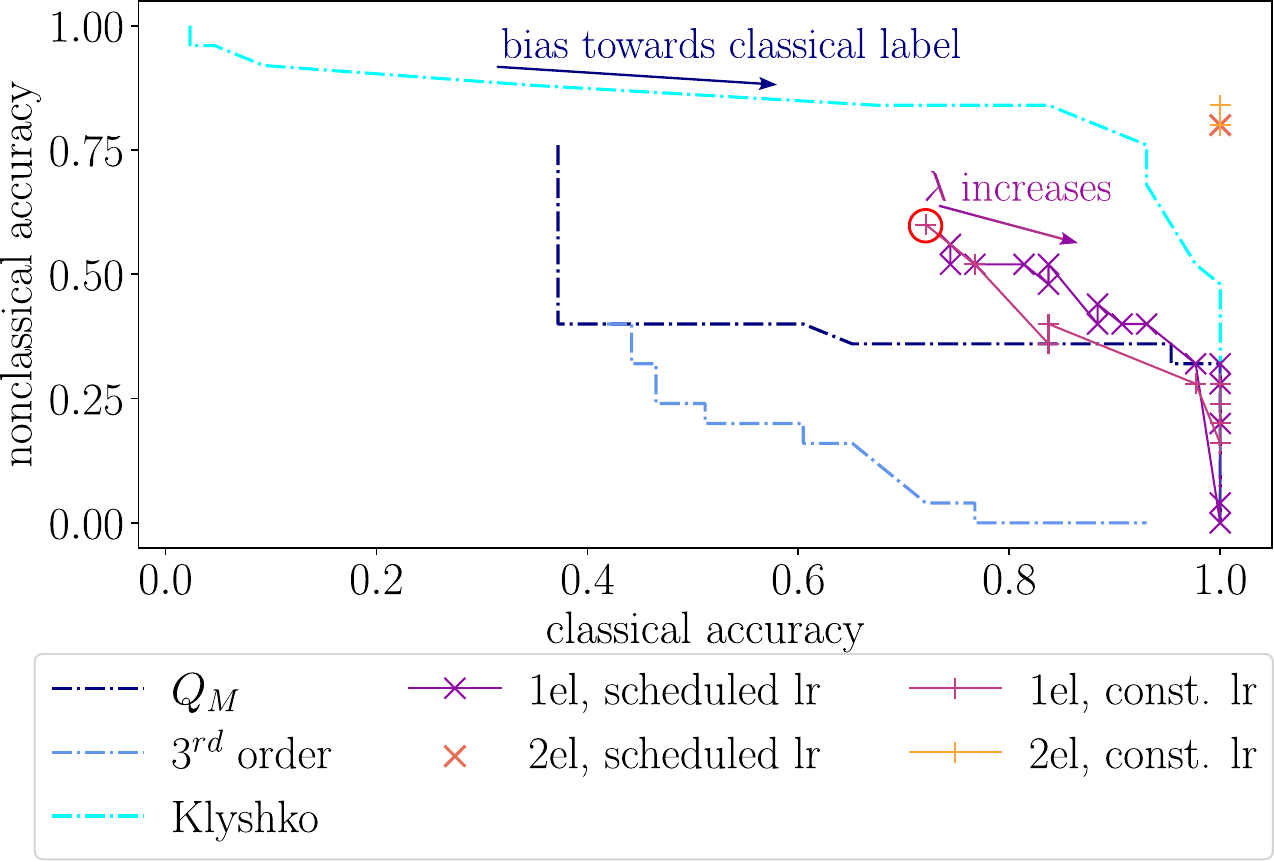}
    \caption{Performance of model on the dataset simulated with a perfect PNR detector and $10^3$ measurements per state. The dashed-dotted lines represent traditional witnesses: The (moment-based) Mandel $Q$ parameter, its third-order generalization and Klyshko's witness (probability-based). The solid lines show the performance of the model trained with one and two encoding layers, corresponding to second- and third-order witnesses: Pluses represent the model being trained with a constant learning rate of $10^{-2}$; the best epoch was chosen according to the best accuracy during training. Crosses represent the model trained with a scheduler that halved the learning rate once a plateau was reached. To visualize the increase of the regularization strength $\lambda$, that implicitly parametrizes the curve, the markers are connected with solid lines. The red circle show the point of $\lambda=0.8$ where the decision rule~\eqref{eq: learned decision rule} is extracted.}
    \label{fig: Performance ds12}
\end{figure}
In order to monitor the quality of this compromise curve, we plot the accuracy in classifying nonclassical states against the accuracy for classical states, as shown in Fig.~\ref{fig: Performance ds12}. The ideal scenario, with $100\%$ accuracy for both classes of states, would correspond to a point in the upper right corner of the plot. Hence, the further a curve penetrates into the upper right quadrant, the better the total accuracy and hence the classifier's performance.

In the figure, the dashed-dotted lines represent traditional witnesses: The Mandel $Q$ parameter, its third-order generalization and Klyshko's witness all of which are parameterized by a bias added to the respective witness.
The solid lines, representing the optimal witness found by the AlCla, are implicitly parameterized by the regularization strength $\lambda$. Indeed, the regularization allows for an almost monotonic increase of the classical accuracy.

Let us first consider a model with a single encoding layer --- learning a second-order criterion --- and a training scheme with a constant learning rate (burgundy pluses in the figure). The model exhibits a comparable performance to the Mandel $Q$ parameter. In fact, the model learns a modification of the Mandel $Q$ parameter, with the same sign structure and approximately equal amplitudes of the coefficients: For a regularization strength of $\lambda=0.8$ (marked as red circle), the found decision rule reads
\begin{equation}
    \begin{split}
    y &= 0.6377\langle \hat{n}^2\rangle-0.5947\langle \hat{n}\rangle^2-0.5754 \langle \hat{n}\rangle-0.3801 \\
    &=0.6377\langle (\Delta\hat{n})^2\rangle -0.5754 \langle \hat{n}\rangle+ 0.0430 \langle\hat{n}\rangle^2  -0.3801.
    \label{eq: learned decision rule}
    \end{split}
\end{equation}
It is noteworthy, that this was the sole instance in which the model learned a physically known witness. In the majority of cases, the model outperforms traditional witnesses by learning a decision rule that cannot be related to some witness derived from the matrix of moments.

When training the model with a scheduler that halves the learning rate once a plateau in the train accuracy is reached, the model outperforms the Mandel $Q$ parameter, as depicted by the violet crosses in Fig.~\ref{fig: Performance ds12}. This indicates that the optimization problem of finding the best nonclassicality indicator is hard and further improvements are possible by using more sophisticated training methods.

For both training schemes, the model is unable to correctly identify more than $73\%$ of the nonclassical states, as some nonclassical features cannot be captured in the space spanned by $\langle \hat{n}\rangle$ and $\langle \hat{n}^2\rangle$. 
In contrast to this, Klyshko's witness takes into account all probabilities $p_0,...,p_{29}$, therefore accessing more fine-grained features of the photon-number distribution than the AlCla which only takes into account up to second-order moments. Consequently, it should come as no surprise that Klyshko's witness reaches a better overall performance.

However, increasing the number of encoding layers to two --- such that the AlCla learns a third-order criterion ---, allows to outperform Klyshko's witness independent of the training method (see orange symbols in Fig.~\ref{fig: Performance ds12}). This demonstrates that a third-order moment-based criterion can be advantageous compared to a probability-based witness, even though the latter takes into account more fine-grained information.
Since the model already achieves an accuracy of $100\%$ on the classical states for a regularization of $\lambda=0$, the curve becomes a vertical line or a single point for training with a constant or an adaptable learning rate, respectively.

\subsection{Learning nonclassicality indicators for experimentally realistic detectors}
In the following, we consider a dataset with two different detection schemes whose POVMs were obtained experimentally: A PNR detector with finite resolution and a time-bin multiplexing scheme. 
To generate measurement samples for a given set of states, we obtain the probabilities of each measurement outcome (number of detected photons/clicks) by multiplying the true number state populations with a matrix representing the detector's POVM. We then generate the datasets by drawing random samples from the resulting low-dimensional probability distributions over observable photon numbers/click counts.

\subsubsection{Finite resolution photon-number detector}\label{sec: results for finite PNR detector}
The first dataset was obtained with the experimentally reconstructed POVM of a finite-resolution photon-number detector. We employed a superconducting nanowire single-photon detector (SNSPD), using the arrival time information relative to a trigger signal to achieve photon-number resolution \cite{sauer_resolving_2023,schapeler_electrical_2024, schapeler_practical_2025}. This allows to resolve 0,1,2,3 and 4+ photons with the last bin containing all higher photon numbers.
The dataset is composed of experimentally measured coherent states, simulated thermal and squeezed states, and SPATS. Table~\ref{tab: ds14} details the composition of the dataset.
\begin{table}[b]
    \caption{\textbf{Composition of the finite resolution PNR detector dataset.} The coherent states were measured experimentally, the remaining states were simulated using the experimentally reconstructed detector POVM with amplitudes $r,\bar{n},\alpha$ lying equidistantly in the amplitude range. There are $27$ classical states and $22$ nonclassical states.}
    \label{tab: ds14}
    \centering
    \begin{tabular}{ccc}
        \toprule\toprule
        species & amplitude range &  number of states \\\midrule
        squeezed & $r\in[0.1, 1.2]$ & 12 \\
        SPATS & $\bar{n}\in[0.15, 0.42]$ & 10\\
        coherent & $\alpha \in [0,12] $ & 13\\
        thermal & $\bar{n}\in[0.5, 7.0]$ & 14\\
        \toprule\toprule
    \end{tabular}
\end{table}
\begin{figure}
    \centering
    \includegraphics[width=\linewidth]{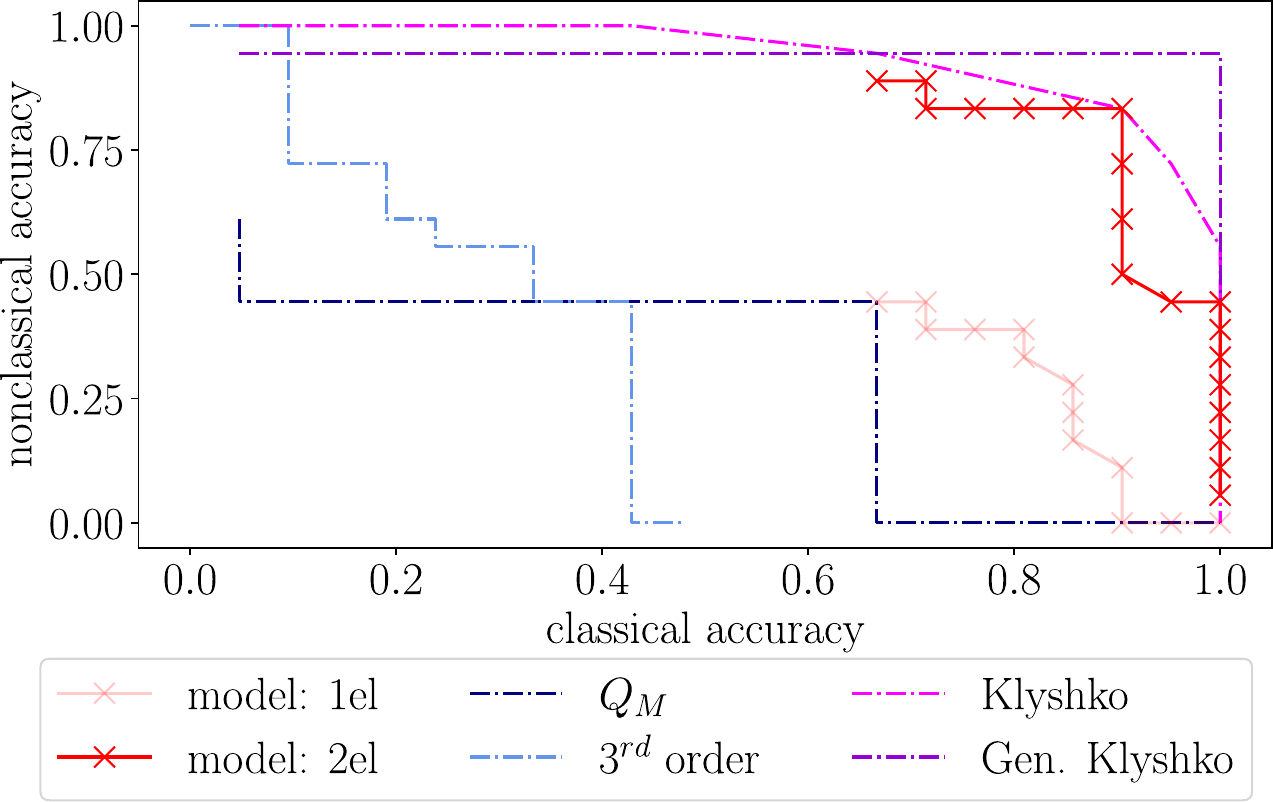}
    \caption{Performance of model on the dataset simulated with a finite resolution photon-number detector and $10^3$ samples per state. The dashed-dotted lines represent traditional nonclassicality witnesses: the Mandel $Q$ parameter, its third-order generalization (both moment-based), Klyshko's witness and the generalized Klyskho witness (both probability-based). The solid red lines correspond to the model being trained with one and two encoding layers, corresponding to a second- and third-order criterion respectively.}
    \label{fig: Performance ds14 1k shots}
\end{figure}
Figure~\ref{fig: Performance ds14 1k shots} shows the performance of the model trained with one and two encoding layers (corresponding to second- and third-order criteria) and $10^3$ samples per state. The model clearly outperforms the Mandel $Q$ parameter and its third-order generalization. 
The reason for this poor performance of traditional moment-based witnesses is the truncation of the photon-number distribution at $n=4$ that leads to a bias towards number squeezing. As a consequence, e.g. coherent states appear number squeezed and are misclassified when naively using the Mandel $Q$ parameter.

On the other hand, Klyshko's and the generalized Klyshko witness prove to be superior, indicating that even for limited statistics, a probability-based approach clearly identifies nonclassicality signatures that are obscured in a moment-based approach due to the truncation.
However, accessing photon-number probabilities becomes increasingly challenging as the number of modes grows (see Sec.~\ref{sec: results 6mode ds}). Furthermore, our finding highlights that traditional witnesses must be carefully adapted to be valid for a given detection scheme, while our approach is completely agnostic of the detection scheme that was used to obtain the training data, showcasing its versatility.

Increasing the number of samples per state to $10^5$ further improves the performance of the probability-based witnesses (not shown in the figure). This is evident because the relative frequencies of photon-counting events become a better proxy for the true probability distribution. Meanwhile, the performance of the AlCla remains unchanged. 

In the process of designing the model, we also tested a dense decoder as a substitute for the algebraic decoder. We found that the decision rule learned by the model with a dense decoder is well approximated by low-order polynomials; furthermore, the dense decoder only yields a moderate improvement in performance compared to the AlCla. Consequently, the simplistic AlCla is sufficiently expressive for the datasets under consideration. A detailed analysis can be found in Appendix~\ref{sec: dense decoder}.

\subsubsection{Time-bin multiplexing scheme} \label{time-bin multiplexing scheme}
The second dataset simulated with an experimentally reconstructed detection POVM was obtained with a multiplexing detection scheme, using two SNSPDs with four time bins each \cite{schapeler_information_2022}. In this scenario, the SNSPDs were used as click detectors with two possible outputs: $0$ and $1$. This corresponds to a total number of $8$ bins in which a photon can trigger a click. The detailed composition of the dataset is given in Tab.~\ref{tab: ds15}.
\begin{table}[b]
    \caption{\textbf{Composition of the time-bin multiplexing dataset.} The coherent states were measured experimentally, the other state species were simulated using the experimentally measured detector POVM with amplitudes $r,\bar{n},\alpha$ lying equidistantly in the amplitude range. There are $27$ classical states and $22$ nonclassical states.}
    \label{tab: ds15}
    \centering
    \begin{tabular}{ccc}
        \toprule\toprule
        species & amplitude range & number of states \\\midrule
        squeezed & $r\in[0.1, 1.2]$ & 12 \\
        SPATS & $\bar{n}\in[0.15, 0.42]$ & 10\\
        coherent & $\alpha \in [1.04\times 10^{-3},9.81\times 10^{1}]$  & 13\\
        thermal & $\bar{n}\in[0.5, 7.0]$ & 14\\
        \toprule\toprule
    \end{tabular}
\end{table}
Due to the nature of the click statistics, the binomial parameter, defined in Eq.~\eqref{eq: def of QB} is considered to be an adequate nonclassicality witness.

The model was trained with one and two encoding layers --- corresponding to second- and third-order criteria --- and $10^3$ measurement samples per state.
\begin{figure}
    \centering
    \includegraphics[width=\linewidth]{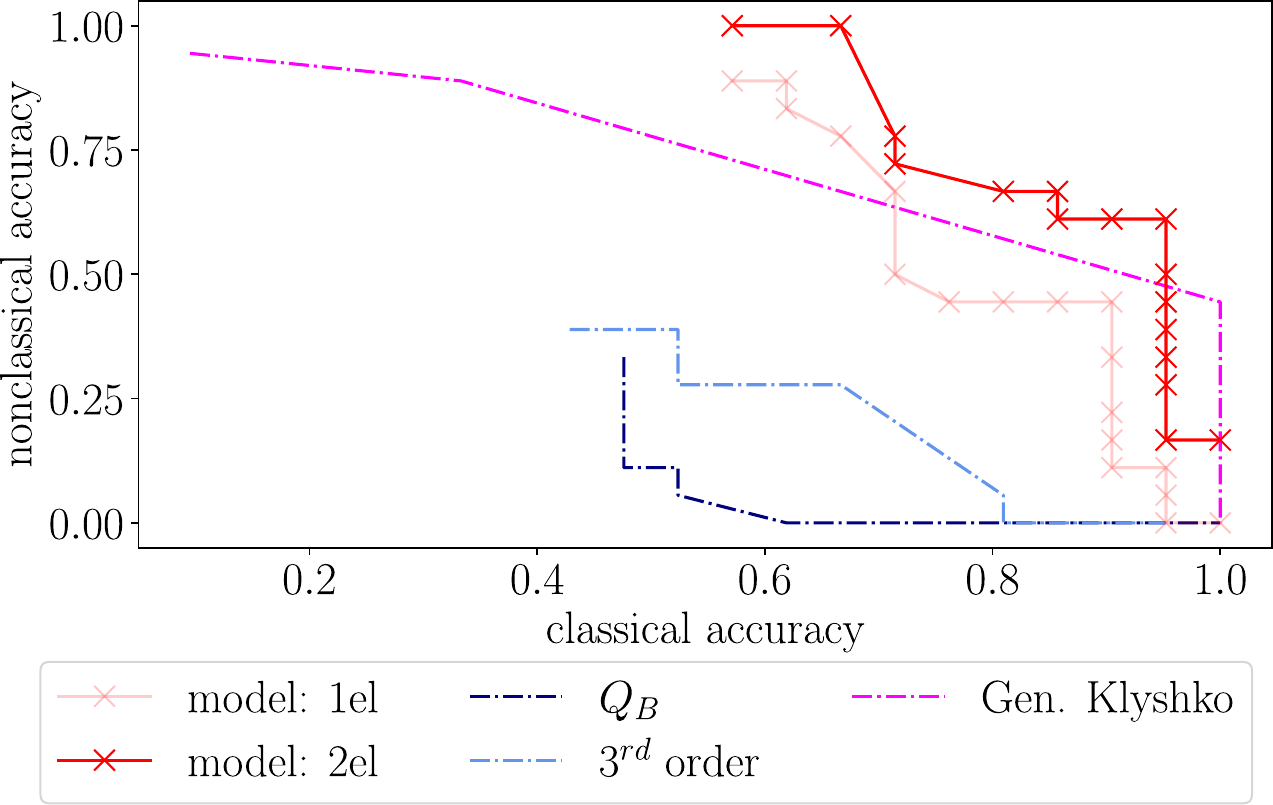}
    \caption{Performance of model on the dataset simulated with a time-bin multiplexing detector and $10^3$ measurements for each state. The dashed-dotted lines represent traditional witnesses: The (moment-based) binomial parameter $Q_B$, its third-order generalization and the (probability-based) generalized Klyshko criterion. The solid red lines represent the model's performance being trained with one or two encoding layers, corresponding to a second- and third-order criterion, respectively.}
    \label{fig: Performance ds15 1kshots}
\end{figure}
Figure \ref{fig: Performance ds15 1kshots} demonstrates that the model outperforms both the binomial parameter $Q_B$ and its third-order generalization. Furthermore, the learned decision rule finds a better compromise than the probability-based generalized Klyshko witness. Increasing the number of samples per state to $9\times10^4$ results in a smaller difference between the generalized Klyshko criterion and the model learning a third-order criterion, as both exhibit comparable performance (not shown in the figure).
The reason for this equivalence may be the nature of the click statistics. Since the click probabilities are defined as a sum over all moments of the photon-number distribution, it is evident that the click probabilities fluctuate, in particular when evaluated with finite statistics. Therefore, it is wise to analyze moments that cumulate the information contained in the full probability distribution. While higher-order moments provide a more complete picture of the probability distribution, there is no guarantee that a high-order witness performs better than a low-order one. This is consistent with the observation that witnesses based on low-order moments, such as the Mandel $Q$ parameter, are often sufficient to prove the nonclassicality of a state \cite{stefszky_benchmarking_2025}.

\subsection{Characterizing a multimode system} \label{sec: results 6mode ds}
We conclude the analysis of the model by demonstrating its generalizability on a 6-mode dataset. The measurement samples were simulated by applying a randomly chosen Gaussian unitary transformation to different sets of product states (coherent, squeezed, and single-photon-loss photon-number states (PNS)), as realized in boson sampling experiments. Photon-number measurements are simulated with the experimentally reconstructed POVM of the finite-resolution photon-number detector described in Sec.~\ref{sec: results for finite PNR detector}. Each state is represented with $M=10^3$ measurement samples; details about the unitary can be found in Appendix~\ref{sec: Appendix 6-mode ds}. Table~\ref{tab: ds16} gives a detailed composition of the dataset.

To visualize the data in a lower-dimensional space, we performed a principal component analysis (PCA). Although the first two principal components explain more than $99\%$ of the variance in the data, the classical and nonclassical states are not fully separable in this projection. Hence, the classification task must be solved in a higher-dimensional space, i.e.\ our more involved approach is necessary. A detailed discussion can be found in Appendix~\ref{sec: Appendix 6-mode ds}.

\subsubsection{Comparison against traditional witnesses}
We benchmark the AlCla against the generalized Klyshko criterion, as well as a 6-mode version of the second-order matrix of moments \cite{agarwal_nonclassical_1992,shchukin_nonclassicality_2005}:
\begin{equation}
    M = \begin{pmatrix}
        1 & \langle:\hat{n}_6:\rangle & ... & \langle:\hat{n}_1:\rangle \\
        \langle:\hat{n}_6:\rangle & \langle: \hat{n}_6^2:\rangle & & \langle:\hat{n}_1\hat{n}_6:\rangle\\
        \vdots& & \ddots & \\
        \langle:\hat{n}_1:\rangle & \langle:\hat{n}_6\hat{n}_1:\rangle &... &\langle:\hat{n}_1^2:\rangle
    \end{pmatrix} \overset{\mathrm{cl.}}{\geq}0.
    \label{def: second-order 6-mode witness}
\end{equation}
A negative minimal eigenvalue of the matrix $M$ verifies the nonclassicality of the state.

\begin{figure}
    \centering
    \includegraphics[width=\linewidth]{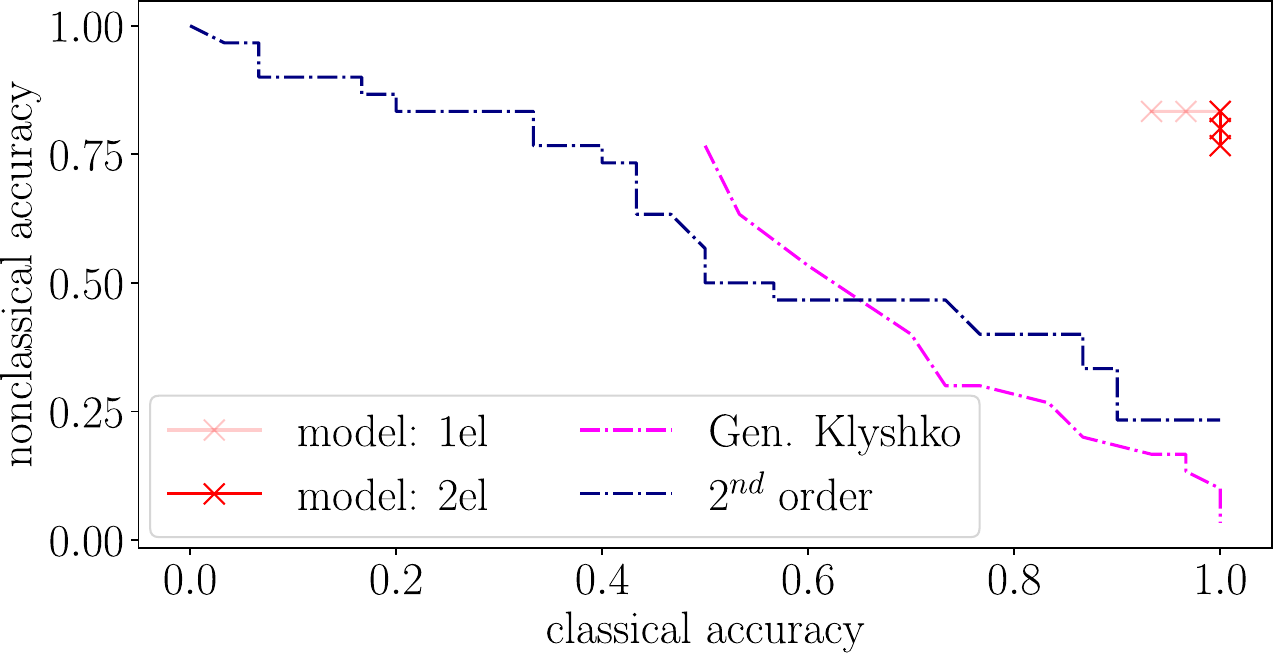}
    \caption{Performance of model on the 6-mode dataset simulated with a fixed random unitary. For each state $10^3$ measurements were taken. The dashed-dotted lines represent the probability-based generalized Klyshko witness and the moment-based second-order witness, defined in Eq.~\eqref{def: second-order 6-mode witness}; the solid red lines represent the model, trained with one and two encoding layers, corresponding to second- and third-order criteria, respectively.}
    \label{fig: Performance ds16 1k shots}
\end{figure}
The model is trained with upper triangular $\lbrace K^{(i)}\rbrace_{i=1,2}$ matrices in the encoder to avoid redundant higher-order moments in the decision rule. 
Figure \ref{fig: Performance ds16 1k shots} shows that the model performs better than the moment-based second-order witness defined in Eq.~\eqref{def: second-order 6-mode witness}. Whereas this witness takes into account all possible two-mode correlations, the AlCla learns to focus on the relevant correlations.

Furthermore, the model clearly outperforms the probability-based generalized Klyshko witness. The reason is that the sparsity of data makes the relative frequency of a photon-counting event an inaccurate estimate for its true probability.

\subsubsection{Enforcing a sparse decision rule}
Training the model on a multimode dataset bears the risk of learning a complicated loss function, relying on all possible correlations between the modes. However, it is possible to enforce a sparse decision rule by adding the additional regularization term $\lambda_K \sum_i^{\#(\textnormal{encoding layers})}\lVert K^{(i)}\rVert_1$ to the loss function.
\begin{figure}
    \centering
    \includegraphics[width=\linewidth]{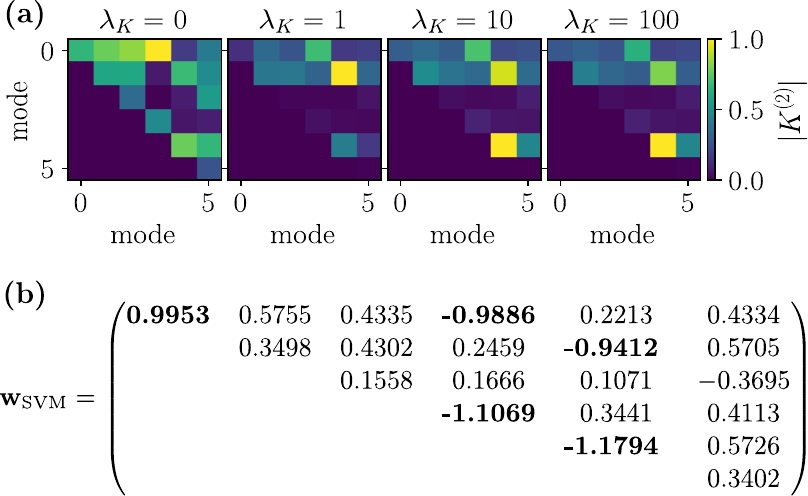}
    \caption{(a) Absolute values of the normalized $K^{(2)}$ encoder weights for a model trained with no regularization $\lambda=0$, one encoding layer and different $K$-regularization strengths $\lambda_K$. (b) Coefficient learned by a linear SVM, reshaped in the form of the matrix $K^{(2)}$. Values with a magnitude larger than $0.94$ are marked bold.}
    \label{fig: ds16 K-Regularization 1kshots}
\end{figure}
 Figure~\ref{fig: ds16 K-Regularization 1kshots} (a) depicts the absolute value of the learned encoder weights $K^{(2)}$, normalized by the maximal absolute value of the matrix. The model was trained with a single encoding layer, a regularization of $\lambda=0$ and different values for the $K$-regularization $\lambda_K$. Increasing the $K$-regularization results in the model shifting its attention from $\langle \hat{n}_0\hat{n}_3 \rangle$ to $\langle\hat{n}_1\hat{n}_4\rangle$ and $\langle \hat{n}^2_4\rangle$.
 \begin{table}[b]
    \caption{\textbf{Accuracy of the model} trained on the 6-mode dataset, with one encoding layer, no regularization $\lambda=0$ and different values for the $K$-regularization $\lambda_K$.}
     \label{tab: ds16 accuracy K_regularization}
     \centering
     \begin{tabular}{c|cc}
     \toprule\toprule
            & \multicolumn{2}{c}{accuracy} \\
         $\lambda_K$ & classical & nonclassical \\\midrule
         0 & 0.9333 & 0.8667\\
         1 & 0.8333& 0.7667\\
         10& 0.8333& 0.7667\\
         100& 0.8333& 0.7667\\
     \toprule\toprule
     \end{tabular}
 \end{table}
 Interestingly, ignoring correlations of the form $\langle \hat{n}_2\hat{n}_{i\geq2}\rangle$ and $\langle \hat{n}_3\hat{n}_{i\geq3}\rangle$ causes the accuracy to drop, but stabilize, as can be seen in Table~\ref{tab: ds16 accuracy K_regularization}. This suggests that it is sufficient to analyze moments associated with the two left-most detectors and the fourth detector. Note that which elements of the correlation matrix are most informative depends on the chosen unitary.

\subsubsection{Comparison against a linear SVM}
To further analyze the model's learned decision rule, we train a support vector machine (SVM) with a linear kernel on the dataset. For each state, all moments $\langle\hat{n}_1\rangle,...,\langle\hat{n}_5\hat{n}_6\rangle,\langle\hat{n}_6^2\rangle$ are computed and flattened to a one-dimensional array $\vec{v}_i\in\mathbb{R}^{27}, i=1,...,N_S$ with $N_S$ being the total number of states in the training dataset.
A linear SVM aims to find a hyperplane in this space that separates differently labeled points from each other. The predicted label is given by $\mathrm{sign}(\vec{w}^T\vec{v}_i+b)$ and the objective is to find $\vec{w}\in\mathbb{R}^{27}$ and $b\in\mathbb{R}$ that maximize the total classification accuracy.
The resulting $\vec{w}$ then describes the normal of the separating hyperplane located at a distance of $|b|/\Vert w\Vert$ from the origin \cite{burges_tutorial_1998}.
On the 6-mode dataset, the linear SVM reaches a total accuracy of $80\%$. For the sake of clarity, we reshape the learned vector $\vec{w}$ into a matrix form that reflects the matrix $K^{(2)}$. Figure~\ref{fig: ds16 K-Regularization 1kshots} (b) shows the learned coefficient $\vec{w}$ with absolute values larger than $0.94$ marked in bold. When comparing $\vec{w}$ with the learned $K^{(2)}$ matrices, we observe a clear correspondence between the increased weights for $\langle \hat{n}_0\hat{n}_3\rangle, \langle \hat{n}_1\hat{n}_4\rangle$ and $\langle \hat{n}_4^2\rangle$ and the respective entries in the $K^{(2)}$ matrix. This finding highlights that the AlCla is capable of extracting the most relevant moments for distinguishing classical from nonclassical states.

\begin{figure}
    \centering
    \includegraphics[width=\linewidth]{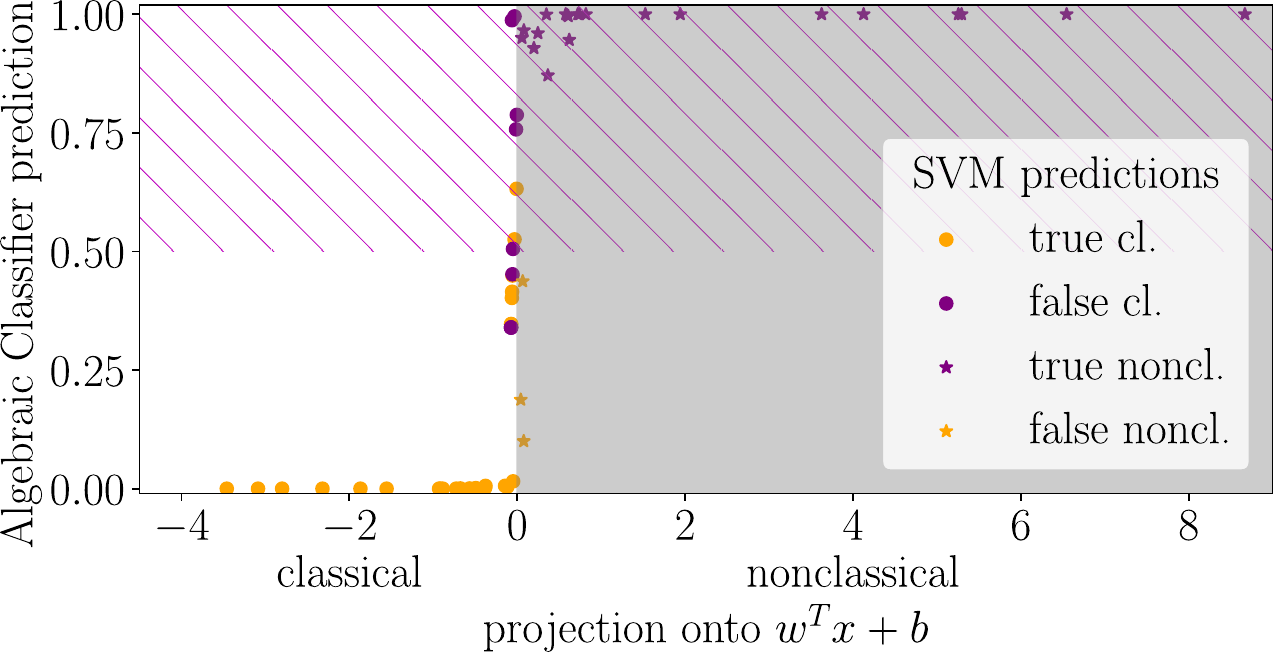}
    \caption{Predictions of the AlCla, trained with one encoding layer and no regularization, compared to the projection $\vec{w}^T\vec{v}+b$ describing the classification of the linear SVM. The hatched (gray shaded) area corresponds to the AlCla's (linear SVM's) prediction that a state is nonclassical. For better visibility, stars represent states identified as nonclassical by the linear SVM, circles represent states identified as classical by the linear SVM. 
    Orange (purple) markers represent classical (nonclassical) states.}
    \label{fig: ds16 SVM}
\end{figure}
Figure \ref{fig: ds16 SVM} compares the predictions of the AlCla with the predictions of the linear SVM. The horizontal axis represents the projection of each state onto $\vec{w}^T\vec{v}+b$ with $\vec{w}$ and $b$ being the coefficient and intercept learned by the linear SVM. 
A value of $\vec{w}^T\vec{v}+b=0$ marks the decision boundary between classical and nonclassical states. For better visibility, states identified as classical (nonclassical) are marked with circles (stars). True classical (nonclassical) states are colored in orange (purple). If the linear SVM could perfectly distinguish between the two classes, the data points would exclusively be orange circles and purple stars.

The vertical axis represents the predictions of the \mbox{AlCla}. Here, a value below (above) 0.5 corresponds to the predicted label "classical" ("nonclassical"). For clarity, the regimes of the AlCla and the linear SVM predicting nonclassicality are hatched in purple and shaded in gray, respectively. The resulting four panels represent the four different cases of the AlCla and the linear SVM giving the same or different predictions.

With respect to the vertical axis, there are few states concentrated around the decision boundary of $y=0.5$ which allows our model to effectively separate the two classes of states. Conversely, several states give a projection $\vec{w}^T\vec{v}+b$ close to zero. This poses a severe challenge for the classification task and explains why our model outperforms the linear SVM.
The key points distinguishing the linear SVM from our model are that (i) the AlCla considers nonlinear terms like $\langle\hat{n}_1\rangle^2$ (therefore is able to rediscover witnesses like the Mandel $Q$ parameter) and (ii) the SVM does not allow for a trimming of the essential features of $\vec{v}$. In fact, the SVM employs all feature dimensions to orient the separating hyperplane in the optimal way. Consequently, the resulting nonclassicality criterion will depend on a multitude of, if not all, potentially computable moments.
 
\section{Conclusions}
\label{sec:conclusions}
We introduced the algebraic classifier (AlCla), a variational model that learns a nonclassicality indicator for photonic multimode quantum states. To our knowledge, this is the first work to present a fully interpretable machine learning-based approach for nonclassicality detection.
By testing the model’s performance against a dense neural network, we demonstrated that the AlCla realizes a unique combination of high expressivity and full interpretability.
The AlCla was trained on data measured with different detection schemes: a perfect PNR detector, a finite-resolution photon-number detector and a time-bin multiplexing click-detection scheme. In all cases, the model outperformed traditional moment-based witnesses. 

Our data-driven approach can be used for any type of measurement data, the only requirement being the availability of a labeled dataset. This versatility contrasts with most traditional nonclassicality witnesses, which have to be carefully adapted to the given detection scheme. Another advantage of our model is its sample efficiency, which is achieved by (i) the use of prior knowledge about sets of relevant state classes, and (ii) the identification and extraction of those photon number/click moments that encode the most significant nonclassical features.

One drawback of our approach is that it requires labeled training data from either a trusted experimental device or a simulation of a noise-free unitary. Additionally, there is no mathematical guarantee that the learned nonclassicality indicator certifies unseen nonclassical states reliably.

Future research could examine nonclassicality criteria formulated in different measurement bases. As our research focused on photon-number and click measurements, a natural next step would be to train on homodyne detection data \cite{gebhart_identifying_2021}. In addition, it is straightforward to implement a hidden dimension into the encoder, which enables parallel training on measurements in multiple different bases. This would enhance the reliability of the learned criterion.
In general, our approach has broad potential as a tool for monitoring the presence of desired resources that are costly to detect experimentally, with possible generalizations, e.g. to the detection of entanglement or non-Gaussianity.

\section*{Acknowledgments}
We thank Anna Dawid and Patrick Emonts for valuable comments. M.J. and M.G. are supported by funding from the German Research Foundation (DFG) under the project identifiers 398816777-SFB 1375 (NOA) and 550495627-FOR 5919 (MLCQS) (supporting also A.B.), from the Carl-Zeiss-Stiftung within the QPhoton Innovation Project MAGICQ, and from the Federal Ministry of Research, Technology and Space (BMFTR) under project BeRyQC. S.K. and J.S. acknowledge the financial support of the Deutsche Forschungsgemeinschaft (DFG) via the TRR 142/3 (Project No. 231447078, Subproject No. C10). T.S. and T.J.B. acknowledge funding by the European Union (ERC, QuESADILLA, 101042399). Views and opinions expressed are however those of the author(s) only and do not necessarily reflect those of the European Union or the European Research Council Executive Agency. Neither the European Union nor the granting authority can be held responsible for them. 
As well as funding from the German Federal Ministry of Research, Technology and Space within the PhoQuant project (grant number 13N16103).

\section*{Data availability and source code}
The code used for this project made use of jax \cite{jax2018github}, flax \cite{flax2020github}, optax \cite{deepmind2020jax}, scikit-learn \cite{scikit-learn} and perceval \cite{heurtel2023perceval}.
We provide the code for training and analyzing the model in the GitHub repository~\url{https://github.com/MartinaJung/IdentifyingOpticalNonclassicality} and training datasets and network parameters in the Zenodo repository~\cite{jung2026learning}. The experimentally measured data can be found in the Zenodo repository~\cite{schapeler2026photon}.


\appendix
\section{Derivation of the moment-based third-order witness}
\label{sec: derivation of 3rd order mom}
In this section, we briefly derive the third-order generalization of the matrix of moments. A more detailed derivation can be found in \cite{krishnaswamy2026observable}.

\subsection{Photon-number-resolving detectors}
Consider the following operator based on the single-mode photon-number operator $\hat{n}$:
$\hat{f} = \sum_{j=0}^\infty f_j \hat{n}^j$. The normal-ordered square of the operator then reads
\begin{equation}
    \langle: \hat{f}^\dagger\hat{f}:\rangle = \sum_{j,k=0}^\infty f_j^*f_k \langle :\hat{n}^{j+k}:\rangle \overset{\mathrm{cl.}}{\geq}0.
\end{equation}
For all classical states, this expectation value must be nonnegative. Hence, the matrix $M_{j,k}$ whose elements are given by $\langle: \hat{n}^{j+k}:\rangle $ must be positive semi-definite. Choosing $j,k\in \mathbb{N} = 0,1,...$ gives the well-known matrix of moments \cite{agarwal_nonclassical_1992, sudarshan_equivalence_1963}:
\begin{equation}
    M = \begin{pmatrix}
        1 & \langle:\hat{n}:\rangle & \langle:\hat{n}^2:\rangle & ...\\
        \langle:\hat{n}:\rangle & \langle:\hat{n}^2:\rangle  & & \\
        \langle:\hat{n}^2:\rangle & & \ddots &\\
        \vdots & & &
    \end{pmatrix}.
\end{equation}
However, when choosing $j,k \in\frac{1}{2}\mathbb{N}\setminus \mathbb{N} = \frac{1}{2},\frac{3}{2},...$, the matrix is given by
\begin{equation}
    M = \begin{pmatrix}
        \langle:\hat{n}:\rangle & \langle:\hat{n}^2:\rangle & \langle:\hat{n}^3:\rangle&...&\\
         \langle:\hat{n}^2:\rangle  &\langle:\hat{n}^3:\rangle & & & \\
        \langle:\hat{n}^3:\rangle & & \ddots &\\
        \vdots & & &
    \end{pmatrix}.
\end{equation}
The first principal minor then leads to the following sufficient third-order criterion for nonclassicality
\begin{equation}
    Q_3 = \langle\hat{n}\rangle\langle:\hat{n}^3:\rangle-\langle:\hat{n}^2:\rangle^2 < 0.
\end{equation}

\subsection{Multiplexing click detection}
For click detection schemes, the derivation follows analogously by choosing 
\begin{equation}
    \hat{f} = \sum_{j=0}^{\lfloor N/2\rfloor}f_j \hat{\pi}^j
    \label{def: operator for click mom}
\end{equation} where $\hat{\pi} = \hat{1}-e^{\frac{\eta\hat{n}}{N} + \nu}$ is the click-POVM with $\eta$ representing the detector efficiency, $\nu$ the dark counts and $N$ the total number of bins in the detection scheme. In comparison to the PNR case, the exponent of the click operator $\hat{\pi}$ is bounded from above by $\lfloor N/2\rfloor$ where $f(x) = \lfloor x \rfloor$ is the floor function that returns the largest integer below the argument $x$. The motivation underlying the upper bound $\lfloor N/2\rfloor$ in the sum becomes evident when considering the normal-ordered expectation value of the operator: For any classical state $\sum_{j,k=0}^{\lfloor N/2\rfloor}f_j^*f_k \langle:\hat{\pi}^{j+k}:\rangle \overset{\mathrm{cl.}}{\geq}0$, hence the matrix of moments reads
\begin{equation}
    M = \begin{pmatrix}
        1 & \langle:\hat{\pi}:\rangle & ...&&\langle:\hat{\pi}^{\left\lfloor \frac{N}{2}\right\rfloor}:\rangle \\
        \langle:\hat{\pi}:\rangle & \langle:\hat{\pi}^2:\rangle  & & \\
        \vdots& & \ddots &\\
        \langle:\hat{\pi}^{\left\lfloor \frac{N}{2}\right\rfloor}:\rangle & & & & \langle:\hat{\pi}^{(2\left\lfloor \frac{N}{2}\right\rfloor)}:\rangle
    \end{pmatrix}\overset{\mathrm{cl.}}{\geq}0.
    \label{eq: def of click-mom even indices}
\end{equation}
The reason for the bound of the exponents in Eq.~\eqref{def: operator for click mom} is the truncation of the counting statistics: Since there only exist $N$ independent probabilities $\lbrace c_k\rbrace_{k=0}^{N-1}$, moments beyond order $N$ must contain redundant information. Therefore, the matrix of moments is finite.

The first principal minor of Eq.~\eqref{eq: def of click-mom even indices} gives the binomial parameter $Q_B = \langle:\hat{\pi}^2:\rangle  - \langle:\hat{\pi}:\rangle^2$. In order to rephrase this expression into a form that depends exclusively on the measurable moments $\langle\hat{c}^m\rangle$, we consider the following generating function
\begin{equation}
    g(x) = \sum_k c_k x^k = \left\langle:\left[ x(1-e^{-\hat{\Gamma}}) + e^{-\hat{\Gamma}} \right]^N :\right\rangle
\end{equation}
where $\hat{\Gamma}= \frac{\eta\hat{n}}{N} + \nu$. Taking derivatives with respect to $x$ and evaluating these at the point $x=1$ gives
\begin{equation}\label{pi_moments rewritten}
\begin{split}
    \langle:\hat{\pi}:\rangle &= \frac{\langle \hat{c}\rangle}{N} = \frac{1}{N}\sum_{k}c_k k\\
    \langle:\hat{\pi}^2:\rangle &= \frac{\langle\hat{c}^2\rangle - \langle\hat{c}\rangle}{N(N-1)} \\
    \langle:\hat{\pi}^3:\rangle &= \frac{\langle\hat{c}^3\rangle-3\langle\hat{c}^2\rangle + 2\langle\hat{c}\rangle}{N(N-1)(N-2)}.
\end{split}
\end{equation}
The first leading principal minor of Eq.~\eqref{eq: def of click-mom even indices} can then be rewritten as $Q_B = \langle\hat{c}^2\rangle - \frac{N-1}{N}\langle\hat{c}\rangle^2 - \langle\hat{c}\rangle$.

When choosing $j,k \in\frac{1}{2}\mathbb{N}\setminus \mathbb{N}$, the matrix of moments reads
\begin{equation}
    M = \begin{pmatrix}
        \langle:\hat{\pi}:\rangle & \langle:\hat{\pi}^2:\rangle &...&\langle:\hat{\pi}^{\left\lceil \frac{N}{2}\right\rceil}:\rangle\\
        \langle:\hat{\pi}^2:\rangle  &\langle:\hat{\pi}^3:\rangle  & \\
        \vdots & & \ddots &\\
        \langle:\hat{\pi}^{\left\lceil \frac{N}{2}\right\rceil}:\rangle& & & \langle:\hat{\pi}^{(2\left\lceil \frac{N}{2}\right\rceil-1)}:\rangle
    \end{pmatrix} \overset{\mathrm{cl.}}{\geq}0.
    \label{eq: def of click-mom odd indices}
\end{equation}
The first leading principal minor yields the sufficient nonclassicality criterion $Q_{B,3} = \langle\hat{\pi}\rangle \langle :\hat{\pi}^3:\rangle - \langle:\hat{\pi}^2:\rangle^2 <0$. Using Eqns.~\eqref{pi_moments rewritten}, we can reformulate the witness in terms of click moments:
\begin{equation}
    Q_{B,3} = \langle \hat{c}^3\rangle\langle \hat{c}\rangle - \frac{N-2}{N-1}\langle\hat{c}^2\rangle^2 - \frac{N+1}{N-1}\langle \hat{c}^2\rangle\langle\hat{c}\rangle + \frac{N}{N-1} \langle \hat{c}\rangle^2.
\end{equation}

\section{Derivation of the generalized Klyshko witness}
\label{sec: derivation of gen. Klyshko}
The derivation of the probability-based witness for photon-number measurement is based on standard methods that have already been used in \cite{filip_hierarchy_2013, innocenti_nonclassicality_2022}.

Consider a detector capable of resolving up to $C+1$ photons. Since the probability $p_{C+1}$ of detecting the maximal possible number of photons is determined by the normalization constraint $\sum_{i=0}^{C+1} p_i = 1$, we only consider the probabilities $p_i : i\in[0,C]$.

\subsection{Single-mode case} The incomplete POVM for photon-number measurements is given by
\begin{equation}
    \hat{\Pi}_j = \; :\frac{\hat{\Gamma^j}}{j!} e^{-\hat{\Gamma}}: \; j\in[0,C]
\end{equation}
with $\hat{\Gamma} = \eta \hat{n}+\nu$ taking into account the detection efficiency $\eta$ and number of dark counts $\nu$. Consider the operator
\begin{equation}
    \hat{f} = \sum_{j=0}^{\lfloor C/2\rfloor} f_j \hat{\Pi}_je^{\hat{\Gamma}/2}
\end{equation}
with arbitrary functions $f_j$. The normal-ordered square of this operator is given by
\begin{equation}
\begin{split}
    :\hat{f}^\dagger\hat{f}: &= \sum_{j,k=0}^{\lfloor C/2\rfloor} f_j^*f_k :\frac{\hat{\Gamma^j}}{j!}e^{-\hat{\Gamma}} \frac{\hat{\Gamma^k}}{k!}:\\
    & = \sum_{j,k}^{\lfloor C/2\rfloor} f_j^*f_k \frac{(j+k)!}{k!j!}: \frac{\hat{\Gamma}^{j+k}}{(j+k)!} e^{-\hat{\Gamma}}:
\end{split}
\end{equation}
Taking the expectation value of this expression and identifying $ \langle:\frac{\hat{\Gamma}^{j+k}}{(j+k)!} e^{-\hat{\Gamma}}:\rangle = p_{j+k}$ yields
\begin{equation}
    0 \overset{\mathrm{cl.}}{\leq} \langle :\hat{f}^\dagger\hat{f}:\rangle = \sum_{j,k}^{\lfloor C/2\rfloor} f_j^*f_k\underbrace{\left[ \binom{j+k}{j} p_{j+k}\right]}_{=:M_{j,k}} \;,\; j+k \leq C.
\end{equation}
Here $p_{j+k}$ is the probability of detecting $j+k$ photons.
In order to fulfill $\langle :\hat{f}^\dagger\hat{f}:\rangle \overset{\mathrm{cl.}}{\geq} 0 \;\; \forall f$, the so-defined matrix $M$ must be positive semi-definite for classical states. Depending on whether $j,k$ are chosen as integers $j,k\in\mathbb{N}$ or half-integers $j,k\in\frac{1}{2}\mathbb{N}\setminus \mathbb{N}$, the witness will be sensitive to odd or even photon-number parity.

Due to the freedom of choice regarding the definition of the operator $\hat{f}$, there exist a plethora of different, yet equivalent, witnesses. In this work, we define the matrix elements with the binomial coefficient to circumvent the fast growth of the factorial that amplifies potential statistical noise in the data.

\subsection{Multimode case} 
The witness can be extended to the multimode scenario. Here, the matrix $M$ remains a 2-dimensional object, while the multimode probabilities are ordered according to a superindex.
Consider a setup with $N$ detectors, each capable of detecting up to $C+1$ photons. The event $(n_0,n_1, ..., n_{N-1})$ can then be mapped to an integer using the unique base-C-to-int mapping:
\begin{itemize}
    \item[(i)] Determine all probabilities $p_{(n_0, n_1, ...n_{N-1})}$ with $n_i \in [0,C]$.
    \item[(ii)] Identify the tuple $(n_1,n_2, ..., n_{N-1})$ as the representation of an integer $\mathtt{c}$ in the basis $C$: $ [n_0,n_1, ..., n_{N-1}]_C = \mathtt{c}$ with 
    $\mathtt{c}= n_0 + n_1\cdot C + ... + n_{N-1} C^{N-1}$.
    \item[(iii)] Define a new probability vector $\vec{p'}$ where $p'(\mathtt{c} =[n_0, n_1, ..., n_{N-1}]_C ) := p(n_0, n_1, ..., n_{N-1})$.
    \item[(iv)] Construct the matrix $M_{j,k}$ using the probability vector $\vec{p'}$.
\end{itemize}

\section{6-mode dataset}
\label{sec: Appendix 6-mode ds}
The 6-mode dataset was simulated with the randomly chosen unitary
\begin{equation}
    \hat{U} = \begin{pmatrix}
        -0.14 &-0.59 & 0.25 &-0.64  &0.23 &-0.32\\
         0.28 & 0.10  &-0.80  &-0.33  &0.40  &-0.00  \\
         0.46 & 0.31 & 0.15 &-0.58 &-0.57 & 0.09\\
        -0.59 & 0.40  & 0.12 &-0.38  &0.24 & 0.53\\
        -0.17 & 0.60  & 0.10  &-0.04  &0.14 &-0.76\\
        -0.56 &-0.15 &-0.50  &-0.07 &-0.62 &-0.17
    \end{pmatrix}
\end{equation}
which can be decomposed exactly into a triangular circuit of beam splitters and phase shifters \cite{reck_experimental_1994, clements_optimal_2016}.
For our purpose of training the model on data from a random optical circuit, it was sufficient to sample the orthogonal group. However, in order to ensure the $\#P$-complexity of the task, one would have to choose a Haar-random unitary.
\begin{table*}[t]
    \caption{\textbf{Composition of the 6-mode dataset.} All measurements were simulated using the experimentally reconstructed POVM of the partially PNR SNSPD with amplitudes $r,\alpha,n$ lying equidistantly in the amplitude range. There are $38$ classical states and $37$ nonclassical states, comprising squeezed and single-photon-loss photon-number states.}
    \label{tab: ds16}
    \centering
    \begin{tabular}{cllc}
        \toprule\toprule
        \multicolumn{2}{c}{input state species} & amplitudes & number of states \\\midrule
        squeezed & $\vec{r}=|r,r,r,r,r,r\rangle$ & $r\in[0.1,0.6]$ & 6 \\
        squeezed & $\vec{r}=|0,0,0,r,r,r\rangle$ & $r\in[0.1,0.8]$ & 8 \\
        squeezed & $\vec{r}=|r,r,r,0,0,0\rangle$ & $r\in[0.1,0.8]$ & 8 \\
        coherent & $\vec{\alpha}=|\alpha, \alpha, \alpha, \alpha, \alpha, \alpha\rangle$ & $\alpha,  \in[0, 0.9]$ & 10\\
        coherent & $\vec{\alpha}=|0,0,0,\alpha, \alpha, \alpha\rangle$ & $\alpha,  \in[0.1,1.4]$  & 14\\
        coherent & $\vec{\alpha}=[\alpha, \alpha, \alpha, 0,0,0]$ & $\alpha,  \in[0.1,1.4]$  & 14\\
        single-photon-loss PNS & $\vec{n} = 0.9|n,n,n,n,n,n\rangle + 0.0167\sum_{\vec{\sigma}} |n-1\rangle_{\sigma_1}\otimes |n\rangle_{\sigma_2} ... \otimes |n\rangle_{\sigma_6}$ & $n\in[1,5]$& 5 \\
        single-photon-loss PNS & $\vec{n} = 0.95|0,0,0,n,n,n\rangle + 0.0167\sum_{\vec{\sigma}}|0,0,0\rangle \otimes |n-1\rangle_{\sigma_1}\otimes|n\rangle_{\sigma_2} \otimes |n\rangle_{\sigma_3}$ & $n\in[1,5]$& 5 \\
        single-photon-loss PNS & $\vec{n} = 0.95|n,n,n,0,0,0\rangle +0.0167\sum_{\vec{\sigma}}|n-1\rangle_{\sigma_1}\otimes|n\rangle_{\sigma_2} \otimes |n\rangle_{\sigma_3}\otimes|0,0,0\rangle $ & $n\in[1,5]$& 5 \\
        \toprule\toprule
    \end{tabular}
\end{table*}
Table~\ref{tab: ds16} gives a detailed overview about the simulated states in the dataset.

\begin{figure}
    \centering
    \includegraphics[width=\linewidth]{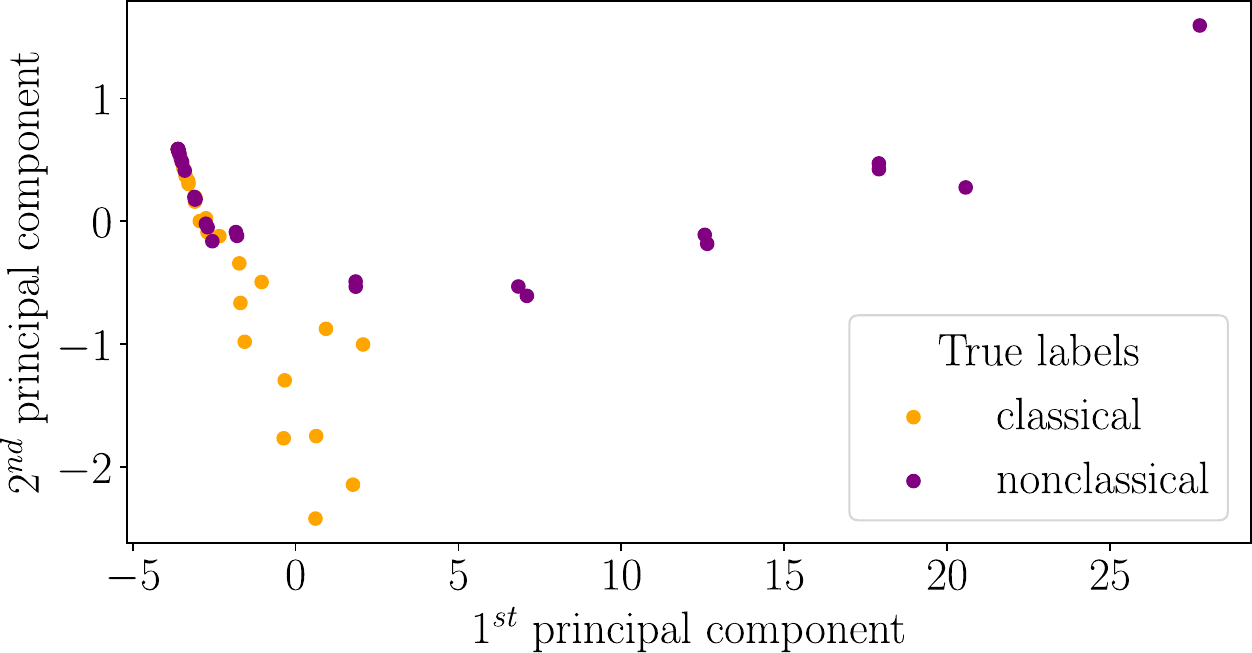}
    \caption{Visualization of the 6-mode dataset with true labels in the space spanned by the first two principal components.}
    \label{fig: PCA ds16}
\end{figure}
Figure~\ref{fig: PCA ds16} visualizes the data points with true labels in the space spanned by the first two principal components, which cumulatively explain $99.60\%$ of the data's variance.
Notably, there is a concentration of both classical and nonclassical data points in the regime of a low first and second principal component. This demonstrates that the assumption underlying PCA --- the largest spread in the data contains most of the information --- does not hold in this case.

\section{Details about the algebraic classifier} \label{sec: details about AlCla}
\subsection{Loss curve}
\begin{figure}
    \centering
    \includegraphics[width=\linewidth]{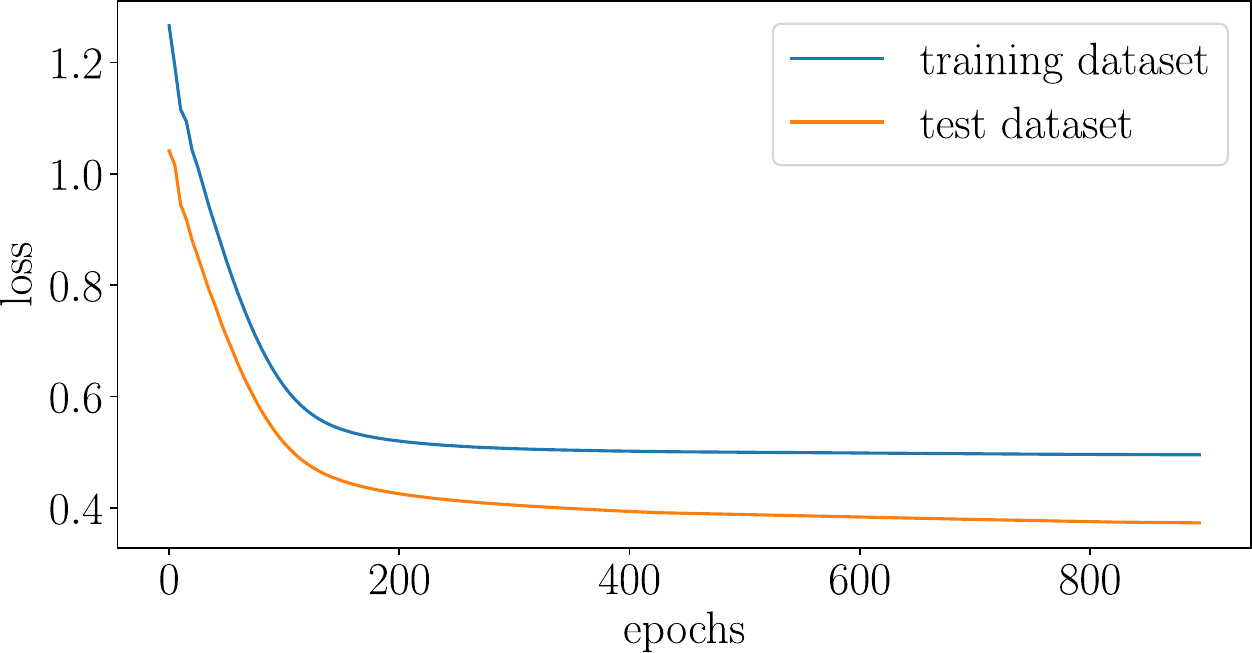}
    \caption{Decay of the loss curve as function of the training epochs for the dataset simulated with the time-bin multiplexing detection scheme and $10^3$ samples per state. The AlCla was trained with one encoding layer and no regularization.}
    \label{fig: typical loss curve}
\end{figure}
Figure~\ref{fig: typical loss curve} shows the decaying value of the loss function on the training and test dataset with the number of epochs. Exemplarily, we chose the single-mode dataset simulated with the time-bin multiplexing detection scheme and $10^3$ samples per state. The AlCla was trained with one encoding layer and no regularization. The extraordinary observation that the test curve lies below the training curve can be explained by the small size of the test dataset (10 states compared to 39 states in the training dataset).

\subsection{Scaling of the decoder parameters}
\begin{figure}
    \centering
    \includegraphics[width=\linewidth]{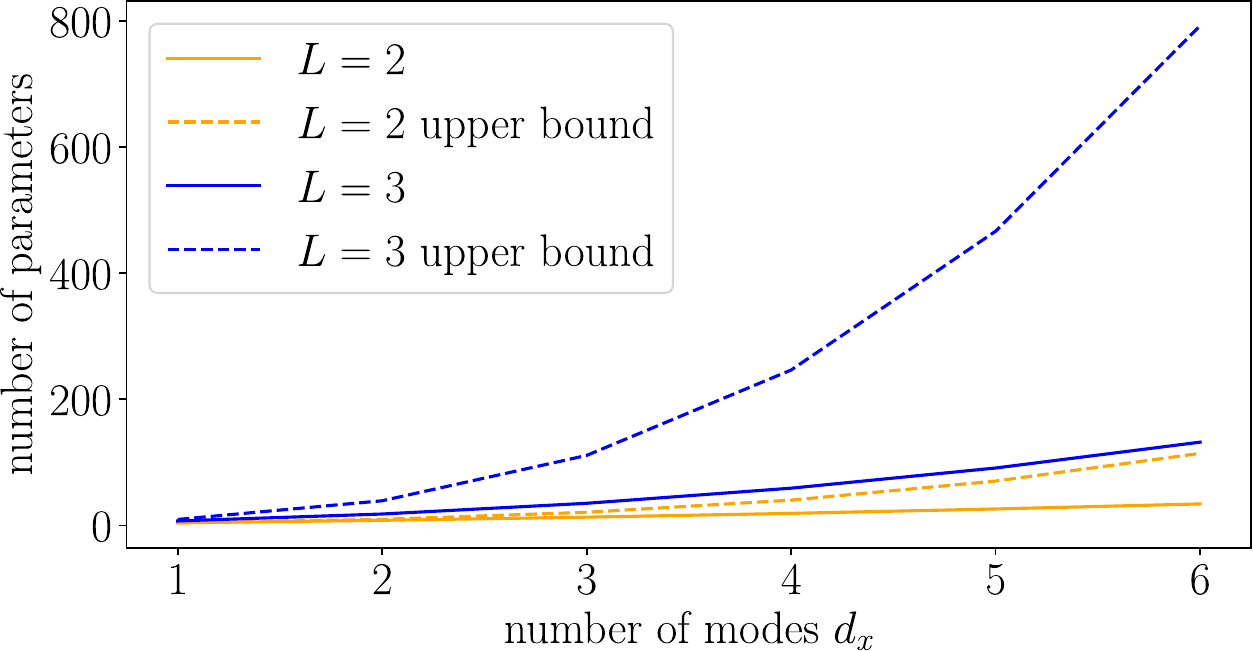}
    \caption{Scaling of the number of parameter in the algebraic decoder with the number of modes $d_x$. Different curves correspond to different orders $L$ of the constructed polynomial. The dashed lines represent the analytical upper bound.}
    \label{fig: scaling of decoder parameters}
\end{figure}
Here we discuss the scaling of the number of decoder parameters in the AlCla with the number of encoding layers $L-1$ and the number of modes $d_x$. General combinatorial arguments lead to an exponential scaling $d_x^L$. However, by further restricting the type of terms considered in the polynomial $f$, this can be reduced significantly. In the following, we consider the case in which the model learns a polynomial that consists only of single, double, and triple terms of the form $\left(x_{i1}^{(m_1)}\right)^{j_1}\left(x_{i2}^{(m_2)}\right)^{j_2}\left(x_{i3}^{(m_3)}\right)^{j_3}$. This results in an upper bound on the parameters count scaling as $(d_xL\log(L))^3$ to leading order in $L$ and $d_x$.
The construction of the decoder polynomial is a combinatorial task, since all $\left\lbrace x_\nu^{(i)} \right\rbrace_\nu^i$ can be combined with each other, given that their cumulative order is at most $L$, i.e. $\sum_i j_im_i\leq L$. For a single mode, the number of parameters $N_P$ in the decoder is given by
\begin{equation}
    \begin{split}
       N_P= &1+\sum_{m=1}^L\left\lfloor\frac{L}{m}\right\rfloor + \frac{1}{2}\sum_{m_1=1}^L\sum_{\substack{m_2=1\\m_1\neq m_2}}^{L-m_1}\sum_{j_1=1}^{\lfloor L/m_1\rfloor}\sum_{j_2=1}^{\lfloor L/m_2\rfloor}1 \\
        +&\frac{1}{3!}\sum_{m_1=1}^L\sum_{\substack{m_2=1\\m_1\neq m_2}}^{L-m_1}\sum_{\substack{m_3=1\\m_1\neq m_3\\
        m_2\neq m_3}}^{L-m_1-m_2}\sum_{j_1=1}^{\lfloor L/m_1\rfloor}\sum_{j_2=1}^{\lfloor L/m_2\rfloor}\sum_{j_3=1}^{\lfloor L/m_3\rfloor} 1\\
        &\textnormal{subject to } \sum_{i}m_ij_i\leq L.
    \end{split}
\end{equation}
First, we lift the constraint $m_i\neq m_j \;\;\forall i\neq j$ and partially include the single and double terms into the triple term. Taking into account that $\left( \left\lfloor \frac{L}{m_i}\right\rfloor\right)^2 \geq \left\lfloor \frac{L}{m_i}\right\rfloor$ since $L\geq 1$, we obtain:
\begin{equation}
    \begin{split}
        N_P\leq &1+\frac{5}{6}\sum_{m=1}^L\left\lfloor \frac{L}{m}\right\rfloor \\
        +& \frac{1}{3!}\sum_{m_1=1}^L\sum_{m_2=1}^{L-m_1}\sum_{m_3=1}^{L-m_1-m_2}
        \sum_{j_1=1}^{\lfloor L/m_1\rfloor}\sum_{j_2=1}^{\lfloor L/m_2\rfloor}\sum_{j_3=1}^{\lfloor L/m_3 \rfloor} 1\\
        &\textnormal{subject to } \sum_{i}m_ij_i\leq L.
    \end{split}
\end{equation}
Next, we lift the constraint $ \sum_{i}m_ij_i\leq L$ and extend the sum over all $m_i$ to $L$ resulting in
\begin{equation}
    \begin{split}
        N_P < 1+\frac{5}{6}\sum_{m=1}^L\left\lfloor \frac{L}{m}\right\rfloor+ \frac{1}{6}\left(\sum_{m=1}^{L}\left\lfloor \frac{L}{m}\right\rfloor\right)^3.
    \end{split}
\end{equation}
Using the harmonic numbers, the summation can be bounded by
\begin{equation}
    \sum_{m=1}^{L}\left\lfloor \frac{L}{m}\right\rfloor \leq L\sum_{m=1}^L\frac{1}{m} = L\left( \ln(L) + \gamma+ \frac{1}{2L}-\epsilon_L\right)
\end{equation}
with $\gamma\approx 0.5572$ the Euler-Mascheroni constant and $\epsilon_L= \frac{1}{8L^2}$. For $L\geq 1$, we have $\sum_{m=1}^{L}\left\lfloor \frac{L}{m}\right\rfloor \leq L\ln(L)+L\gamma=:\tilde{L}$
and $N_P < 1+\frac{\tilde{L}}{6}\left(5+\tilde{L}^2\right)$.
Note that this is not a tight bound but a strong approximation since lifting the constraint $\sum_i m_ij_i\leq L$ allows for terms of the order $L^3$. For higher values of $L$, the bound becomes increasingly loose.

For $d_x$ modes, additional summations $\sum_{i_1}...$ precede the sum and counting becomes more delicate as further combinations with $m_1=m_2, i_1\neq i_2$ become possible. Taking into account that there are $3!$ possibilities to distribute $i_1,i_2,i_3$ over three modes, we get $N_P < 1+\frac{d_x\tilde{L}}{36}\left(35+d_x^2\tilde{L}^2\right)$. Depending on $\tilde{L}$, the first or second term in the bracket will dominate. Figure \ref{fig: scaling of decoder parameters} shows the scaling of $N_P$ with the number of modes $d_x$. For $L=2$, $N_P$ and the analytical upper bound are plotted as solid and dashed orange curves, respectively. Blue curves describe the case $L=3$. As expected, the analytical upper bound is not tight. Hence, the number of parameters in the decoder scales polynomially, yet not faster than $d_x^3$.

\section{Dense decoder}
\label{sec: dense decoder}
\subsection{Architecture}
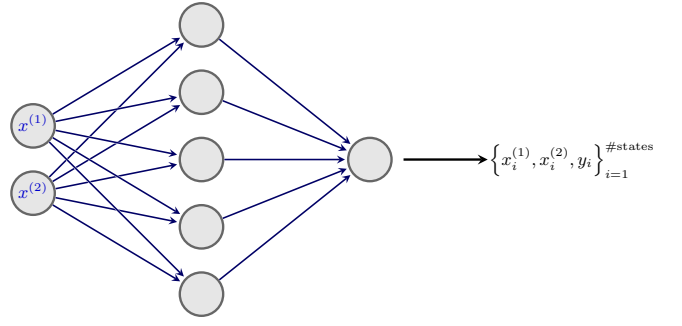
\begin{figure}
\centering
\resizebox{\columnwidth}{!}{%
\begin{tikzpicture}[x=3.0cm,y=1.2cm]

  \readlist\Nnod{2,5,1} 
  \readlist\Nstr{2,5,1} 
  \readlist\Cstr{x,h,y} 
  \def\yshift{0.} 
  
  \foreachitem \N \in \Nnod{
    \def\lay{\Ncnt} 
    \pgfmathsetmacro\prev{int(\Ncnt-1)} 
    \foreach \i [evaluate={\c=int(\i==\N); \y=\N/2-\i-\c*\yshift;
                 \x=\lay; \n=\nstyle;
                 \index=(\i<\N?int(\i):"\Nstr[\n]");}] in {1,...,\N}{ 
    \ifnum \lay<2
        \node[node 1] (N\lay-\i) at (\x,\y) {$\strut\Cstr[\n]^{(\index)}$};
    \fi
    \ifnum \lay>1
        \node[node 1] (N\lay-\i) at (\x,\y) {};
    \fi
    
      \ifnumcomp{\lay}{>}{1}{ 
        \foreach \j in {1,...,\Nnod[\prev]}{ 
          \draw[white,line width=1.2,shorten >=1] (N\prev-\j) -- (N\lay-\i);
          \draw[connect] (N\prev-\j) -- (N\lay-\i);
        }
      }

    }
  }
\draw[-stealth, very thick] (3.2,-0.5) -- (3.7, -0.5);
\node (output) at (4.2, -0.5) {$\left\lbrace x_i^{(1)},x_i^{(2)}, y_i\right\rbrace_{i=1}^{\# \mathrm{states}} $};
\end{tikzpicture}}
\caption{Structure of the dense decoder, implemented for a model with one encoding layer. Input to the decoder are the first $\vec{x}^{(1)}$ and the second order moments $\vec{x}^{(2)}$.}
\label{tikzfig: architecture of dense decoder}
\end{figure}
In order to judge the sufficiency of the simple algebraic classifier, a more sophisticated model was used to learn the optimal indicator of nonclassicality for a given dataset. 
In this approach, the algebraic decoder is replaced by a dense neural network, schematically depicted in Fig.~\ref{tikzfig: architecture of dense decoder}. 
In this architecture, the encoder's output is input to a dense feed-forward neural network whose number of input nodes is equal to $d_xL$ for $d_x$ modes and $L-1$ encoding layers. The encoder outputs are processed by three dense layers with ReLU activation functions in the first two layers.

In Fig.~\ref{fig: Performance ds14 - dense decoder vs approx polynomials}, the solid red line shows the performance of the model with a dense decoder trained with two encoding layers on the single-mode dataset with finite PNR measurements and $10^5$ samples per state. Without conducting an extensive hyperparameter search, we found the best performance with $(10,4,1)$ hidden nodes per layer. The model gives a moderate improvement compared to the AlCla (see Fig.~\ref{fig: Performance ds14 1k shots}, which illustrates the performance on the dataset with $10^3$ samples per state; however, it also quantitatively correctly describes the performance of the model for $10^5$ samples per state). In fact, the model with a dense decoder achieves $100\%$ accuracy for nonclassical states. However, when examining the performance for larger values of the regularization strength, both models realize $100\%$ accuracy for the classical and around $50\%$ accuracy for the nonclassical states. Hence, both models represent faithful nonclassicality indicators, yet the dense decoder provides no advantage over the AlCla.

\subsection{Polynomial regression}
\begin{figure}
    \includegraphics[width=\linewidth]{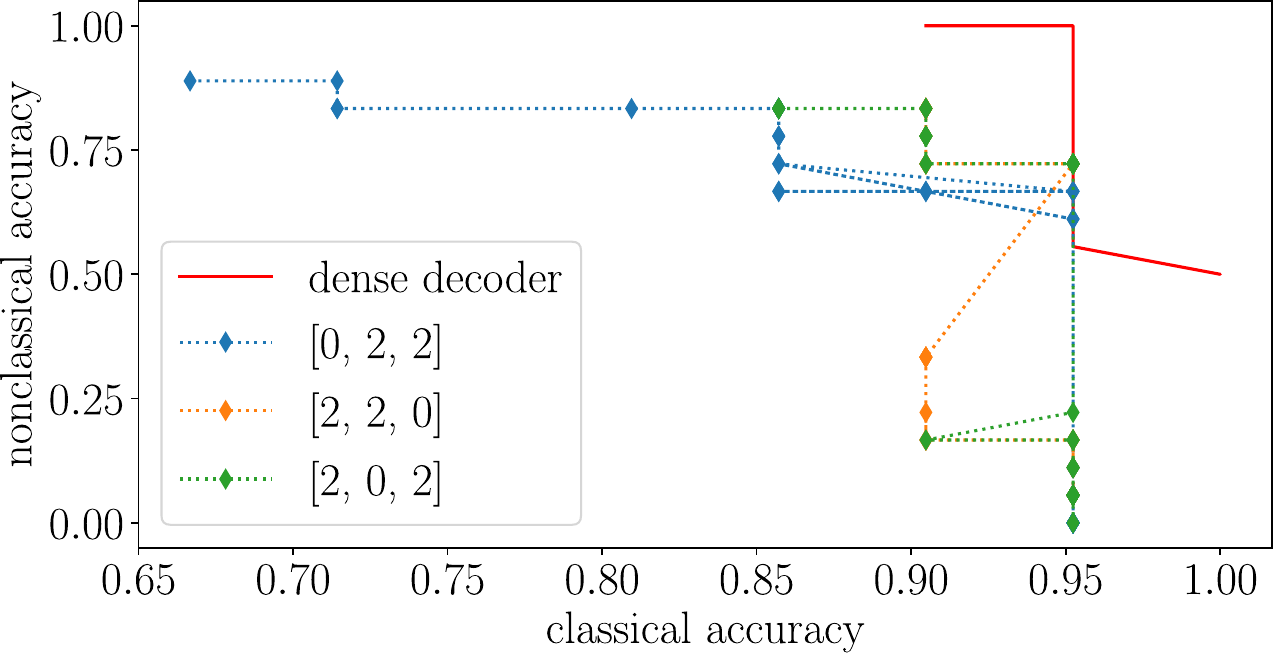}
    \caption{Performance of dense decoder model (red solid line) versus approximating polynomials (diamonds) on the single-mode dataset with finite PNR measurements and $10^5$ samples per state. The model was trained with two encoding layers. Lines connecting the markers represent the increase in the regularization strength $\lambda$.
    The expression in the brackets gives the maximum order of the monomials, i.e.\ [$e_1, e_2, e_3$] corresponds to the highest-order term in the polynomial being $(x^{(1)})^{e_1}, (x^{(2)})^{e_2}, (x^{(3)})^{e_3}$ with $x^{(1)},x^{(2)},x^{(3)}$ the outputs of the encoder.}
    \label{fig: Performance ds14 - dense decoder vs approx polynomials}
\end{figure}
In order to approximate the learned decision rule of the dense decoder, the tuples $\left\lbrace x_i^{(1)}, x_i^{(2)}, x_i^{(3)}, y_i\right\rbrace_{i=1}^{N_S}$, comprising the input and output of the decoder, are fitted with a hierarchy of polynomial regressions. Here, $N_S$ denotes the number of states. The quality of the fit is analyzed with the coefficient of determination
\begin{equation}
    R^{2}(\vec{y},\vec{f}) = 1 - \frac{\sum_{i=1}^{N_S} (y_i - f_i)^2}{\sum_{i=1}^{N_S} (y_i - \bar{y})^2}
\end{equation}
with $f_i$ the predicted value of the $i^{th}$ sample, $y_i$ the true label, and $\bar{y} = \frac{1}{N}\sum_{i=1}^{N_S} y_i$. Training the dense decoder without regularization and two encoding layers resulted in a coefficient of determination well above $93\%$. 
Figure \ref{fig: Performance ds14 - dense decoder vs approx polynomials} compares the performance of the dense decoder model to the approximating polynomials. Here, the tuple $[e_1,e_2,e_3]$ represents the maximal exponent of the encoder outputs $x^{(1)},x^{(2)},x^{(3)}$ in the polynomial. For instance, $[2,2,0]$ corresponds to a function $f$ linear in $x^{(1)},(x^{(1)})^2, x^{(2)},(x^{(2)})^2$. Since the polynomials are fitted to the dense decoder, they naturally do not exhibit a monotonicity with respect to the regularization strength $\lambda$. In terms of overall performance, the polynomials achieve a similar total accuracy as the model with a dense decoder. These findings indicate that low-order polynomials are sufficient for achieving a high accuracy on the dataset.

\bibliography{references.bib}

\end{document}